# Strong coupling expansion for scattering phases in Hamiltonian lattice field theories

## I. The $(d+1)$-dimensional Ising model

Bernd Dahmen

Deutsches Elektronen-Synchrotron DESY
Notkestrasse 85, D-22603 Hamburg, Germany

**Abstract**

A systematic method to obtain strong coupling expansions for scattering quantities in Hamiltonian lattice field theories is presented. I develop the conceptual ideas by means of the Hamiltonian field theory analogue of the Ising model, in $d$ space and one time dimension. The main result is a convergent series representation for the sacttering states and the transition matrix. To be explicit the special cases of $d = 1$ and $d = 3$ spatial dimensions are discussed in detail. I compute the next-to-leading order approximation for the phase shifts. The application of the method to investigate low-energy scattering phenomena in lattice gauge theory and QCD is proposed.

# 1 Introduction

The investigation of scattering phenomena in lattice field theories has attracted increasing interest in recent years. Since the work of Lüscher [1], who established a close relation between the two-particle spectrum in a sufficiently large volume and the elastic scattering amplitude, several authors have succeeded to extract scattering data from numerical simulations of lattice models in 2,3 and 4 space-time dimensions [2]–[5]. As further steps towards the determination of the low-energy scattering parameters of QCD there have recently been encouraging attempts to investigate resonance phenomena [6]–[8], and even some first full QCD calculations have been performed [9, 10].

This paper advocates a different method to study scattering in lattice field theories. In particular I present a systematic technique to obtain strong coupling expansions for the scattering matrix and the phase shifts in the elastic regime, applicable to a large class of lattice models including gauge theory and QCD. Such a proposal may look a bit old-fashioned. But if one remembers that strong coupling expansions have provided important qualitative insight into numerous issues of lattice field theories it is astonishing that only little effort has been made to apply this powerful analytical method for a computation of scattering quantities [11]–[13]. To close this gap towards a thorough understanding of scattering in lattice field theories is the intent my approach.

Since scattering processes involve asymptotic two-particle states whose real time evolution is not directly accessible on a euclidean lattice I prefer the Hamiltonian formulation to develop the basic ideas. For many lattice field theories the Hamiltonian consists of two essentially different terms

$$\mathsf{IH} = \mathsf{IH}_0 + \kappa \mathsf{IH}_1 \ , \tag{1.1}$$

where $\kappa$ is inversely proportional to some coupling constant. The first part, $\mathsf{IH}_0$, is a sum of terms which refer to a single lattice site, whereas $\mathsf{IH}_1$ couples neighbouring sites. As long as the inverse coupling $\kappa$ is zero the system is purely static and it can usually be solved, if not analytically, then numerically. Turning on the parameter $\kappa$ is what makes the dynamics set in. Actually $\mathsf{IH}_1$ creates and annihilates lattice excitations at different sites and thereby induces their propagation through the lattice. One may try to describe the scattering of those excitations by means of a perturbative expansion in the inverse coupling.

As the test case I choose the Hamiltonian field theory analogue of the Ising model, in $d$ space and one time dimension. This theory is obtained by considering the transfer matrix of the Ising model in the extremely anisotropic limit of the exchange couplings [14]. It is complicated enough to confront us with most of the conceptual difficulties of the problem. Still the model is sufficiently simple (no resonances, for example) such that the basic issues are not obscured by technicalities.

The organization of this paper is as follows. To begin with I discuss the static limit of the $(d+1)$-dimensional Ising model. After briefly recalling a systematic perturbation theory for isolated levels this method is applied to the static vacuum, the one-particle and two-particle sector. The outcome are effective Hamiltonians which determine the dynamics of the system in these sectors. The rest of the paper is devoted to the diagonalization of the effective two-particle Hamiltonian. In section 3 I describe the general method to contruct convergent series representations for the scattering states and the transition matrix. The leading and next-to-leading order contributions are calculated explicitly. I continue by studying the simplest case, the $(1+1)$-dimensional system in some detail. In section 4 the next-to-leading order approximation for the scattering states is determined in closed analytical form such that the physical properties can be investigated. Life is substantially more complicated in the case of $d = 3$ spatial dimensions which is considered in section 5. At small momentum the continuum rotational symmetry is restored on the lattice. Thereby the diagonalization of the scattering matrix simplifies considerably . By means of a transformation to better suited momentum variables the analysis can be extended beyond the small momentum regime. The next-to-leading order approximation for the scattering phases is computed. This paper ends with a few concluding remarks and is completed by three appendices, two discussing the physical properties of the one-particle and the scattering states and one discussing the calculation of the three-dimensional Green's function.

# 2 Effective Hamiltonians

The system studied here is the Ising model in $(d+1)$ dimensions. In the static limit where the inverse coupling $\kappa$ is zero the spectrum consists of isolated eigenvalues. This is the prerequisite for a systematic perturbative expansion of the physical quantities by means of a very elegant method developed by Bloch [15]. The leading and next-to-leading order correction to the static vacuum, one-particle and two-particle sector are calculated.

## 2.1 The $(d+1)$-dimensional Ising model and its static limit

We consider a $d$-dimensional hypercubic spatial lattice

$$\Gamma = \{x \mid x/a \in \mathbb{Z}^d, -L/2 < x_k \leq L/2\} \tag{2.2}$$

with linear extent $L$ and spacing $a$ which is chosen equal to 1 for convenience. So $L$ is an integer number and the physical quantities are dimensionless. Furthermore we assume periodic boundary conditions. At each site $x$ of $\Gamma$ we associate a copy of $\mathbb{C}^2$ and call it $\mathbb{C}^2_x$. The unit vectors

$$\mathrm{e}_1(x) = \begin{pmatrix} 1 \\ 0 \end{pmatrix} \quad , \quad \mathrm{e}_0(x) = \begin{pmatrix} 0 \\ 1 \end{pmatrix} \tag{2.3}$$

are referred to as spin up and spin down states at $x$, respectively. For finitely many spins a Hilbert space $\mathcal{H}$ is defined as the tensor product of all $\mathbb{C}^2_x$. Let $\sigma_1, \sigma_2, \sigma_3$ denote the familiar Pauli matrices and $\sigma_i(x)$ represent the operators on $\mathcal{H}$ that act by $\sigma_i$ on the $x$ component of the tensor product and by the identity on the others. These operators commute at different sites which guarantees that the corresponding spins are independent observable quantities.

The field theory version of the Ising model, in $d$ space and one (continuous) time dimension, has the Hamiltonian [14]

$$\mathsf{IH}' = \frac{g}{2}\{\mathsf{IH}_0 + \kappa \mathsf{IH}_1\} \ , \tag{2.4}$$

where

$$\mathsf{IH}_0 = \sum_{x\in\Gamma}\{\mathbb{1} - \sigma_3(x)\} \ , \tag{2.5}$$

$$\mathsf{IH}_1 = -\sum_{x\in\Gamma}\sum_{k=1}^{d}\sigma_1(x)\sigma_1(x+\hat{k}) \ . \tag{2.6}$$

The parameter $g$ is a dimensionless coupling constant, $\kappa = 2/g^2$ and $\hat{k}$ denotes the unit vector in the positive $k$-direction. For the following it is convenient to define

$$\mathsf{IH} = \frac{2}{g}\mathsf{IH}' = \mathsf{IH}_0 + \kappa\mathsf{IH}_1 \ . \tag{2.7}$$

As already mentioned in the introduction the Hamiltonian consists of two essentially different terms. The static part $\mathsf{IH}_0$ does not couple different lattice sites. On the contrary, $\sigma_1$ is the operator that flips an up spin down and a down spin up. So the hopping terms $\sigma_1(x)\sigma_1(x+\hat{k})$ flip the spins on neighbouring sites and thereby propagate signals through the lattice.

The idea of this work is to investigate scattering phenomena by means of a perturbative expansion in the inverse coupling $\kappa$. To begin with we have to determine the spectrum of the unperturbed static Hamiltonian $\mathsf{IH}_0$. The $\kappa = 0$ ground state $|\Omega\rangle^0$ is determined through the condition

$$\mathsf{IH}_0|\Omega\rangle^0 = 0 \ , \tag{2.8}$$

i.e. the strong-coupling vacuum is the unique state having all spins up. It is normalized such that ${}^0\langle\Omega|\Omega\rangle^0 = 1$. We define

$$|x\rangle^0 = \sigma_1(x)|\Omega\rangle^0 \tag{2.9}$$

as the state where the spin is down at site $x$ and all others are up. One finds

$$\mathrm{IH}_0|x\rangle^0 = 2|x\rangle^0 \ , \tag{2.10}$$

hence $\sigma_1(x)$ creates a one-particle state of energy 2. The successive action of $\sigma_1$ leads to multi-particle configurations

$$|x^1,\ldots,x^n\rangle^0 = \sigma_1(x^1)\ldots\sigma_1(x^n)|\Omega\rangle^0 \tag{2.11}$$

which are the higher excitations of the static system

$$\mathrm{IH}_0|x^1,\ldots,x^n\rangle^0 = 2n|x^1,\ldots,x^n\rangle^0 \ . \tag{2.12}$$

Since the spin-flip operators at different sites commute, any permutation of $\{x^1,\ldots,x^n\}$ in the above definition (2.11) represents the *same* $n$-particle state. But it has to be emphazised that $\sigma_1$ and $\sigma_1^\dagger = \sigma_1$ are no bosonic creation and annihilation operators in the usual sense. Rather they are unitary operators satisfying the constraint $\sigma_1^2 = \mathbb{1}$ which leads to $\sigma_1(x)\sigma_1(x)|\Omega\rangle^0 = |\Omega\rangle^0$. That is, two excitations cannot occupy the same lattice site and the multi-particle states are characterized by a set of mutually different vectors in $\mathbb{Z}^d$. So the $n$-particle eigenspaces of $\mathrm{IH}_0$ are given by

$$\mathcal{E}_n = \mathrm{span}\left\{|x^1,\ldots,x^n\rangle^0 \mid x^1 \neq \ldots \neq x^n, x^k \in \Gamma\right\} \ . \tag{2.13}$$

They are well separated by an energy gap of 2 and highly degenerate.

## 2.2 Bloch's perturbation theory

We are concerned with the self-adjoint operator $\mathrm{IH} = \mathrm{IH}_0 + \kappa\,\mathrm{IH}_1$ defined on the Hilbert space $\mathcal{H}$. The static part $\mathrm{IH}_0$ has a purely discrete spectrum with eigenvalues $2n$. The projection operators on the corresponding eigenspaces $\mathcal{E}_n$ are denoted by $\mathrm{P}_n$.

If the perturbation $\kappa\mathrm{IH}_1$ is turned on, one expects that each static eigenspace $\mathcal{E}_n$ is deformed into a linear space $\mathcal{H}_n \subset \mathcal{H}$. In $\mathcal{H}_n$ there is a complete (discrete or continuous) set of states $|n,\alpha\rangle$ such that

$$\mathrm{IH}|n,\alpha\rangle = E_n(\alpha)|n,\alpha\rangle \ , \quad E_n(\alpha) = 2n + \epsilon_n(\alpha) \ , \quad \epsilon_n(\alpha) \ = \ \mathcal{O}(\kappa) \ . \tag{2.14}$$

To obtain the eigenstates $|n,\alpha\rangle$ and the eigenvalues $\epsilon_n(\alpha)$ as an explicit power series of the inverse coupling Bloch proceeds in two steps. First the Hamiltonian $\mathrm{IH}$ is reduced to a hermitian operator $\mathrm{H}'_n$ acting on $\mathcal{E}_n$ whose eigenvalues are equal to $\epsilon_n(\alpha)$ to *all* orders of $\kappa$. In a second step this effective Hamiltonian is diagonalized. Let $|n,\alpha\rangle^0$ denote the eigenvectors corresponding to the energy values $\epsilon_n(\alpha)$. Then the announced expansion of the eigenvectors of $\mathrm{IH}$ is obtained by means of a linear operator $\mathrm{U}_n$ which maps $|n,\alpha\rangle^0$ onto $|n,\alpha\rangle$.

Both $\mathrm{H}'_n$ and $\mathrm{U}_n$ are explicitly constructed in Bloch's paper [15]. If the inverse coupling is sufficiently small they are well-defined power series of $\kappa$. Each summand is a product of the perturbation $\mathrm{IH}_1$ and the projection operators $\mathrm{P}_n$ and $\mathrm{Q}_n = \mathbb{1} - \mathrm{P}_n$. In this work we only need the leading and next-to-leading order terms of the hermitean reduced Hamiltonian

$$\mathrm{H}'_n = \kappa\mathrm{P}_n\mathrm{IH}_1\mathrm{P}_n + \kappa^2\mathrm{P}_n\mathrm{IH}_1\frac{\mathrm{Q}_n}{2n-\mathrm{IH}_0}\mathrm{IH}_1\mathrm{P}_n + \mathcal{O}(\kappa^3) \ . \tag{2.15}$$

and the linear operator

$$\mathrm{U}_n = \mathrm{P}_n + \kappa\frac{\mathrm{Q}_n}{2n-\mathrm{IH}_0}\mathrm{IH}_1\mathrm{P}_n + \mathcal{O}(\kappa^2) \ . \tag{2.16}$$

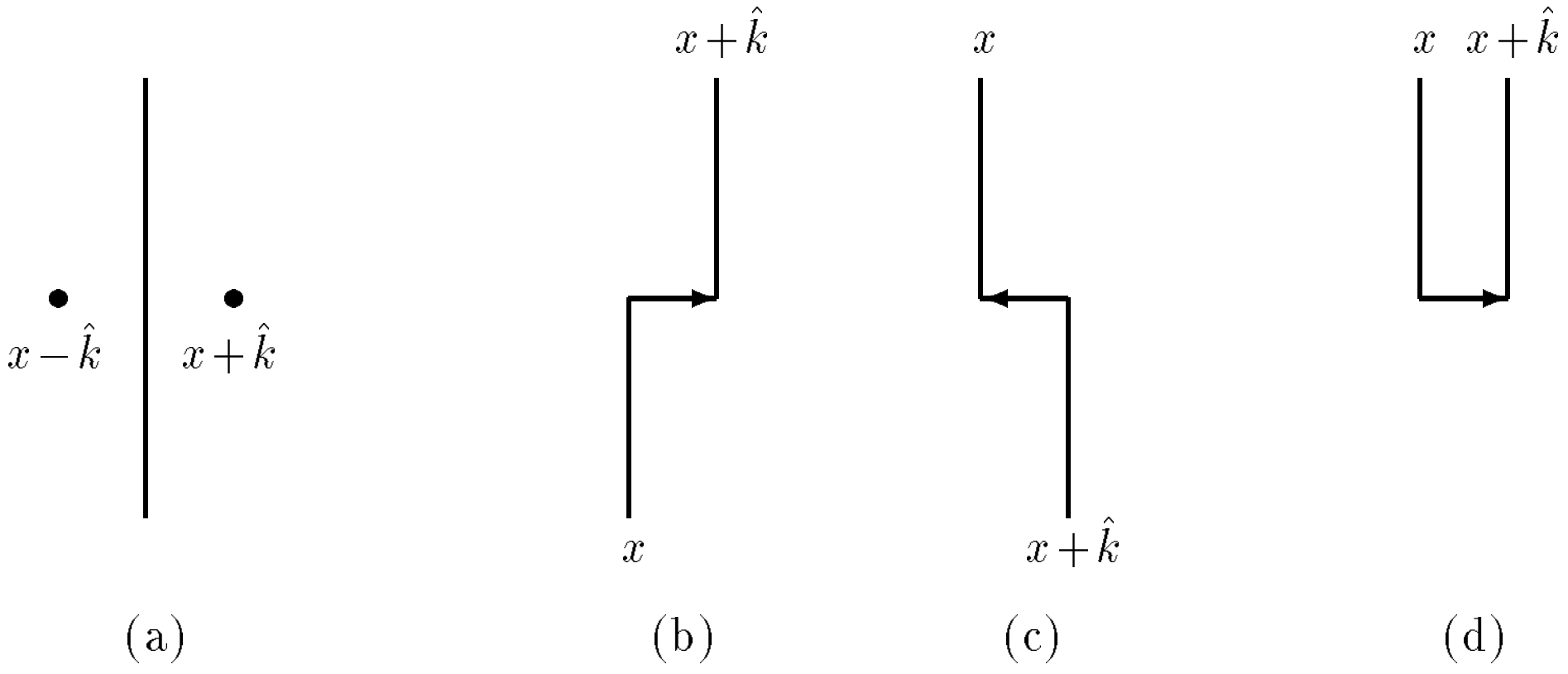

Figure 1: Elementary graphs.

### 2.3 The vacuum sector

The vacuum eigenspace $\mathcal{E}_0$ is the one-dimensional subspace of $\mathcal{H}$ with projection operator $\mathrm{P}_0 = |\Omega\rangle^{0\,0}\langle\Omega|$. The action of the perturbation

$$\mathrm{IH}_1 = -\sum_{k=1}^{d}\sum_{x}\sigma_1(x)\sigma_1(x+\hat{k}) \tag{2.17}$$

on the strong-coupling vacuum $|\Omega\rangle^0$ creates two-particle excitations with flipped spins at neighbouring sites. From (2.15) it follows that the leading order contribution to the reduced Hamiltonian $\mathrm{H}_0'$ vanishes since states from different static eigenspaces are matched. In the next-to-leading order term $\mathrm{Q}_0$ represents the sum over all intermediate states orthogonal to $\mathcal{E}_0$. But here only the two-particle states contribute for the same reason. Hence $\mathrm{Q}_0$ can be replaced by $\mathrm{P}_2$, the projection operator on $\mathcal{E}_2$, and we get

$$\mathrm{H}_0' = -\frac{\kappa^2}{4}\,{}^0\langle\Omega|\mathrm{IH}_1\mathrm{P}_2\mathrm{IH}_1|\Omega\rangle^0\,|\Omega\rangle^{0\,0}\langle\Omega| + \mathcal{O}(\kappa^3)\ . \tag{2.18}$$

It turns out that it is convenient to have a pictorial description of the processes in strong coupling perturbation theory. To evaluate the matrix elements one has to determine the action of the perturbation on the initial state on the r.h.s. and match the result with the final state on the l.h.s. In figs. 1 the sucessive process of this evaluation is visualized by going along the vertical axis from the bottom to the top. Each vertical line represents a state where the spin is flipped down at a certain lattice site $x \in \mathbb{Z}^d$. If it is not affected by the perturbation the configuation remains unchanged. The corresponding diagram is depicted in fig. 1 (a). The hopping term $\sigma_1(x)\sigma_1(x+\hat{k})$ flips the spins on adjacent sites, indicated by a horizontal line in fig. 1 (b) – (d). If it acts on an initial state where the spin is down at site $x$ this spin is flipped up and his neighbour in the positive $k$-direction is flipped down. Correspondingly the signal is moved along the negative $k$-direction in case that the initial state has a spin down at site $x+\hat{k}$. Finally, if both at $x$ and $x+\hat{k}$ the spins are up initially they are flipped down by the action of $\sigma_1(x)\sigma_1(x+\hat{k})$. Following these rules the graph contributing to (2.18) is easily found and depicted in fig. 2. The action of $\mathrm{IH}_1$ on the initial vacuum creates an intermediate two-particle state, denoted by a dashed horizontal line. The spins are flipped back again by the subsequent action of the perturbation and we end up with the final vacuum state. Because there are $dL^d$ pairs of adjacent sites throughout the lattice, the graph carries a counting factor of $dL^d$. We find

$$\mathrm{H}_0' = -\kappa^2\frac{d}{4}L^d|\Omega\rangle^{0\,0}\langle\Omega| + \mathcal{O}(\kappa^3)\ , \tag{2.19}$$

and the ground-state energy is

$$E_0 = -\kappa^2\frac{d}{4}L^d + \mathcal{O}(\kappa^3)\ . \tag{2.20}$$

Though $E_0$ diverges as $L \to \infty$, this does not cause any trouble as long as the energy difference between the ground-state and the excited states stays finite. For later use we construct the next-to-leading order approximation of the ground state by means of Bloch's linear operator $\mathrm{U}_0$ (cf. eq. 2.16)

$$|\Omega\rangle = \mathrm{U}_0|\Omega\rangle^0 = |\Omega\rangle^0 + \frac{\kappa}{4}\sum_{k=1}^{d}\sum_{x}|x, x+\hat{k}\rangle^0 + \mathcal{O}(\kappa^2)\ . \tag{2.21}$$

So when $\kappa$ is small but not zero the vacuum begins to mix with nearest-neighbour two-particle excitations.

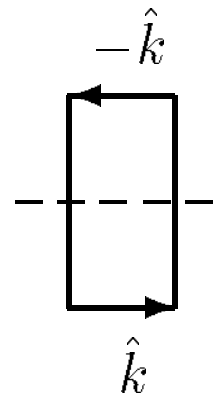

Figure 2: Next-to-leading order contribution to the vacuum energy.

### 2.4 The one-particle sector

Next we consider the eigenspace $\mathcal{E}_1$ spanned by the one-particle excitations $\{|x\rangle^0 \mid x \in \Gamma\}$ of energy 2. The projection operator is

$$\mathrm{P}_1 = \sum_x |x\rangle^{0\,0}\langle x|\ , \tag{2.22}$$

and the leading order hermitean reduced Hamiltonian reads

$$\mathrm{H}_1'^{(1)} = \sum_{x^1,x^2} {}^0\langle x^1|\mathsf{H}_1|x^2\rangle^0|x^1\rangle^{0\,0}\langle x^2|\ . \tag{2.23}$$

The non-vanishing contributions are described by the elementary graphs figs. 1 (b) and (c). The initial flipped spin is moved one site in the positive or negative $k$-direction, respectively, hence

$$\mathrm{H}_1'^{(1)} = -\sum_{k=1}^{d}\sum_x\left(|x\rangle^{0\,0}\langle x+\hat{k}| + |x+\hat{k}\rangle^{0\,0}\langle x|\right)\ . \tag{2.24}$$

In next order of perturbation theory the intermediate configurations are three-particle excitations of energy 6. Here the first contribution is made up of those graphs where the initial one-particle state $|x^1\rangle^0$ remains undisturbed while a vacuum fluctuation occurs on a distant pair of sites. This situation is shown in fig. 3 (a). Because there are $d(L^d-2)$ distant sites the counting-factor is $d(L^d-2)$. The second contributions come from those graphs where the flipped spin of the initial state is shifted over two lattice sites leading to the final states $|x^2\rangle^0 = |x^1 \pm \hat{k} \pm \hat{l}\rangle^0$ and $|x^2\rangle^0 = |x^1 \pm \hat{k} \mp \hat{l}\rangle^0$. They are represented by the graphs fig. 3 (b) – (e) with counting factor 1. We obtain

$$\begin{aligned}\mathrm{H}_1'^{(2)} \;&=\; -\frac{d}{4}(L^d-4)\mathrm{P}_1 - \frac{1}{4}\sum_x\sum_{k,l=1}^{d}\Big(|x\rangle^{0\,0}\langle x+\hat{k}+\hat{l}| + |x+\hat{k}+\hat{l}\rangle^{0\,0}\langle x| \\ &\quad +|x\rangle^{0\,0}\langle x+\hat{k}-\hat{l}| + |x+\hat{k}-\hat{l}\rangle^{0\,0}\langle x|\Big)\ .\end{aligned} \tag{2.25}$$

The additional contribution to the first term is due to the fact that for the graphs fig. 3 (d) and (e) the case $\hat{k} = \hat{l}$ is excluded. For later convenience we prefer to write the summation over the directions without restriction and therefore have to add $\frac{1}{2}d\,\mathrm{P}_1$ as a correction.

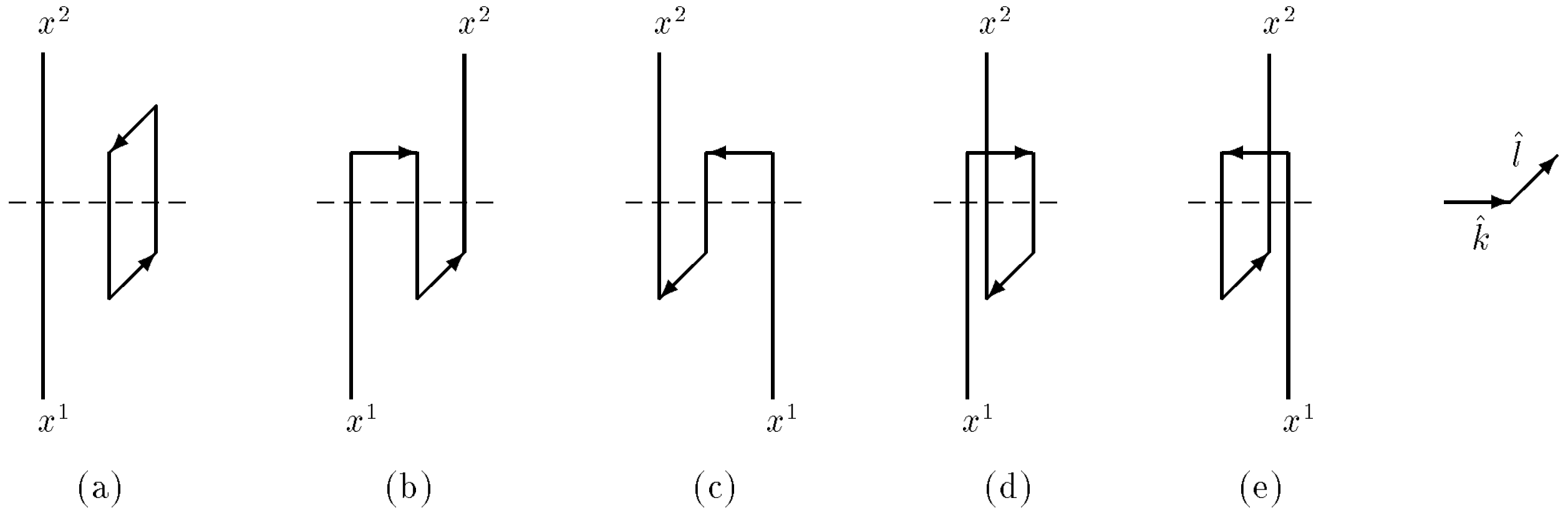

Figure 3: Next-to-leading order graphs with initial state $|x^1\rangle^0$ and final state $|x^2\rangle^0$.

The next step within Bloch's programme is the diagonalization of the reduced Hamiltonian

$$\mathrm{H}_1' = \kappa \mathrm{H}_1'^{(1)} + \kappa^2 \mathrm{H}_1'^{(2)} + \mathcal{O}(\kappa^3) \ . \tag{2.26}$$

This is achieved straightforwardly if we pass to the coordinate space. Any state $|\psi\rangle^0 \in \mathcal{E}_1$ has a unique expansion

$$|\psi\rangle^0 = \sum_x \psi(x)|x\rangle^0 \ , \tag{2.27}$$

where $\psi(x)$ is a complex lattice function. The coordinate space representation of the Hamiltonian (2.26) is determined as follows

$$(\mathrm{H}_1'\psi)(x) = {}^0\langle x|\mathrm{H}_1'|\psi\rangle^0 \ . \tag{2.28}$$

To make the result more readable it is convenient to introduce forward and backward difference operators

$$(\nabla_k\psi)(x) \stackrel{\mathrm{def}}{=} \psi(x+\hat{k}) - \psi(x) \ , \quad (\nabla_k^*\psi)(x) \stackrel{\mathrm{def}}{=} \psi(x) - \psi(x-\hat{k}) \ , \tag{2.29}$$

and the lattice Laplacian

$$\Delta \stackrel{\mathrm{def}}{=} -\sum_{k=1}^{d} \nabla_k^* \nabla_k \ . \tag{2.30}$$

In terms of these we find

$$\mathrm{H}_1' = \kappa\{\Delta - 2d\} - \kappa^2 \left\{ \frac{1}{4}\Delta^2 - d\Delta + d(d-1) + \frac{d}{4}L^d \right\} + \mathcal{O}(\kappa^3) \ . \tag{2.31}$$

One recognizes that the vacuum energy (2.20) contributes to (2.31). Subtracting it from $\mathrm{H}_1'$ we define the resulting operator as a new effective one-particle Hamiltonian which no longer depends on the linear extent $L$ of the lattice

$$\mathrm{H}_1 = \mathrm{H}_1' - E_0 \ . \tag{2.32}$$

Consequently, the infinite volume limit can be taken trivially and we can extend our analysis to the infinite lattice $\Gamma = \mathbb{Z}^d$. In this case the static one-particle sector is represented by the Hilbert space of complex wave functions on $\mathbb{Z}^d$ satisfying

$$||\psi||^2 \stackrel{\mathrm{def}}{=} \sum_x |\psi(x)|^2 < \infty \ . \tag{2.33}$$

Because $\mathrm{H}_1$ commutes with space translations by integer vectors $(\mathrm{U}_v\psi)(x) = \psi(x+v)$, its eigenvectors are plane waves $\exp(ipx)$. The momentum $p$ ranges over the first Brillouin zone $\mathcal{B} = \{p \mid -\pi < p_k \leq \pi\}$. Using the abbreviations

$$\widehat{p}^2 \stackrel{\mathrm{def}}{=} \sum_{k=1}^{d} \widehat{p}_k\ \widehat{p}_k \ , \quad \widehat{p}_k \stackrel{\mathrm{def}}{=} 2\sin\left(\frac{p_k}{2}\right)$$

the eigenvalues read

$$\epsilon_1(p) = \kappa\,(\hat{p}^2 - 2d) - \frac{\kappa^2}{4}\left\{(\hat{p}^2)^2 - 4d\hat{p}^2 + 4d(d-1)\right\} + \mathcal{O}(\kappa^3)\ . \tag{2.34}$$

The physical rest mass of the one-particle excitation is defined as the zero momentum energy

$$m = E_1(0) = 2 - 2d\kappa - \kappa^2 d(d-1) + \mathcal{O}(\kappa^3)\ . \tag{2.35}$$

At some critical inverse coupling $\kappa = \kappa_c$ (2.35) vanishes and the Ising model undergoes a continuous phase transition. In the following we will always assume that $\kappa < \kappa_c$. In this case the one-particle energy is a positive, smooth periodic function of the momentum $p$. The kinetic mass $m_*$ of the particle is defined through the small momentum expansion

$$E_1(p) = m + \frac{p^2}{2m_*} + \mathcal{O}(p^4)\ . \tag{2.36}$$

In the present case one obtains

$$m_*^{-1} = 2\kappa(1 + d\kappa) + \mathcal{O}(\kappa^3)\ . \tag{2.37}$$

At strong coupling the kinetic mass is much larger than the rest mass of the particle. This precisely reflects the fact that the one-particle configurations do not like to move, and in the limit $\kappa \to 0$, they are just static objects sitting on the sites of the lattice.

To check the reliability of our method we discuss the physical properties of the resulting one-particle states. The translationally invariant static (improper) one-particle states read

$$|p\rangle^0 = (2\pi)^{-\frac{d}{2}} \sum_x \mathrm{e}^{\mathrm{i}px} |x\rangle^0\ . \tag{2.38}$$

They are normalized such that ${}^0\langle p|q\rangle^0 = \delta(p-q)$ with $\delta(p)$ denoting the ordinary ($d$-dimensional) Dirac $\delta$-function. The next-to-leading order approximation of the true (improper) one-particle states is constructed as

$$|p\rangle = \mathrm{U}_1|p\rangle^0 = |p\rangle^0 + \kappa\frac{(2\pi)^{-\frac{d}{2}}}{4}\sum_{k=1}^{d}{\sum_{x,y}}' \mathrm{e}^{\mathrm{i}px}|x, y, y+\hat{k}\rangle^0 + \mathcal{O}(\kappa^2)\ . \tag{2.39}$$

Here the prime indicates the restriction $x \neq y$ and $x \neq y+\hat{k}$ such that the summation is taken over three-particle configurations only. Below this notation is always assumed, i.e. in primed sums over static multi-particle states coinciding arguments are excluded. Normalizable one-particle states can then be formed by smearing with any smooth function $\tilde{f}(p)$ which is periodic in all momentum components with period $2\pi$

$$|f\rangle = \int_{\mathcal{B}} \mathrm{d}^d p \tilde{f}(p)|p\rangle = (2\pi)^{\frac{d}{2}}\sum_x f(x)|x\rangle^0 + \kappa\frac{(2\pi)^{\frac{d}{2}}}{4}\sum_{k=1}^{d}{\sum_{x,y}}' f(x)|x, y, y+\hat{k}\rangle^0 + \mathcal{O}(\kappa^2)\ . \tag{2.40}$$

Since the Fourier transform $f(x)$ of $\tilde{f}(p)$ falls off rapidly as $|x| \to \infty$, one intuitively expects that the particle in this state is (essentially) localized in a bounded region $\mathcal{R}$ of space which depends on the wave function $\tilde{f}(p)$. The precise expression of this physical property is as follows (cf. ref. [16]). Let $\mathbb{O}(x)$ be a local operator representing an 'observable' of the theory and assume that $x$ is far outside the region $\mathcal{R}$. A measurement of this operator in the state $|f\rangle$ yields the value $\langle f|\mathbb{O}(x)|f\rangle$ in the average. Now if the particle is essentially confined to $\mathcal{R}$ the measurement of $\mathbb{O}(x)$ should give the same result as in the vacuum state $|\Omega\rangle$. Hence the one-particle state $|f\rangle \in \mathcal{H}$ is well localized if

$$\langle f|\mathbb{O}(x)|f\rangle \longrightarrow \langle f|f\rangle\langle\Omega|\mathbb{O}(x)|\Omega\rangle \tag{2.41}$$

for $|x| \to \infty$ and any local observable $\mathbb{O}(x)$.

In appendix A we show that this condition is indeed satisfied. So we have good reason to argue that (2.40) represent the next-to-leading order approximation of the well localized one-particle states in the $(d+1)$-dimensional Ising model.

## 2.5 The two-particle sector

For the determination of the two-particle scattering states we have to apply perturbation theory to the subspace of static two-particle excitations of energy 4 with projection operator

$$\mathrm{P}_2 = \frac{1}{2}\sum_{x,y}{}' |x,y\rangle^{0\,0}\langle y,x| \ . \tag{2.42}$$

To leading order the reduced two-particle Hamiltonian is given by

$$\mathrm{H}_2'^{(1)} = \frac{1}{4}\sum_{x^1,y^1}{}' \sum_{x^2,y^2}{}' \ {}^0\langle y^1,x^1|\mathrm{IH}_1|x^2,y^2\rangle^0 |x^1,y^1\rangle^{0\,0}\langle y^2,x^2| \ . \tag{2.43}$$

The perturbation acts on the initial state $|x^1,y^1\rangle^0$ by moving one of its flipped spins as described by the graphs in figs. 1 (b) and (c). The other excitation is not affected as represented by the elementary graph in fig. 1 (a). Their union are the graphs contributing to (2.43). Each of them carries a counting factor of 4 because a permutation $x^1 \leftrightarrow y^1$ in the initial state or $x^2 \leftrightarrow y^2$ in the final state leads to the same contribution. We find

$$\mathrm{H}_2'^{(1)} = -\sum_{x,y}{}' \sum_{k=1}^{d} \Big( |x,y\rangle^{0\,0}\langle y+\hat{k},x| + |x,y+\hat{k}\rangle^{0\,0}\langle y,x| \Big) \ . \tag{2.44}$$

The first contribution to the next-to-leading order correction is described by the graph in fig. 4 (a). The initial state consists of two excitations on adjacent lattice sites $y^1 = x^1 \pm \hat{k}$, so the action of $\mathrm{IH}_1$ can flip both spins back. The final state is created from the vacuum at an arbitrarily shifted position. If the creation of the final state occurs before the initial state is annihilated this leads to an intermediate four-particle excitation depicted in fig. 4 (b). The permutation symmetry is accounted for by a counting factor of 4. Further contributions come from vacuum fluctuations (fig. 4 (c)) where the initial state remains unchanged. In case that its excitations are located on neigbouring sites the counting factor is $2(dL^d - 4d + 1)$. Otherwise, the number of distant pairs of sites is decreased by 1 and the counting factor is $2(dL^d - 4d)$. Finally, the perturbation can shift one of the initial excitations over two lattice sites. The corresponding graphs are those of figs. 3 (b) – (e) completed by fig. 1 (a). Again each of them carries a counting factor of 4. We get

$$\begin{aligned}
\mathrm{H}_2'^{(2)} \ = \ & -\frac{d}{4}(L^d-8)\mathrm{P}_2 - \sum_x \sum_{k=1}^{d} |x,x+\hat{k}\rangle^{0\,0}\langle x+\hat{k},x| + \frac{1}{2}\sum_x \sum_{k,l=1}^{d} \Big( |x,x+\hat{k}\rangle^{0\,0}\langle x+\hat{l},x| \\
& + |x,x-\hat{k}\rangle^{0\,0}\langle x+\hat{l},x| + |x,x+\hat{k}\rangle^{0\,0}\langle x-\hat{l},x| + |x,x-\hat{k}\rangle^{0\,0}\langle x-\hat{l},x| \Big) \\
& - \frac{1}{4}\sum_{x,y}{}' \sum_{k,l=1}^{d} \Big( |x,y\rangle^{0\,0}\langle y+\hat{k}+\hat{l},x| + |x,y+\hat{k}+\hat{l}\rangle^{0\,0}\langle y,x| \\
& + |x,y\rangle^{0\,0}\langle y+\hat{k}-\hat{l},x| + |x,y+\hat{k}-\hat{l}\rangle^{0\,0}\langle y,x| \Big) \ .
\end{aligned} \tag{2.45}$$

Since $\mathrm{P}_2$ is just the identity operator on $\mathcal{E}_2$ the infinite volume Hamiltonian is given by

$$\mathrm{H}_2' = \kappa \mathrm{H}_2'^{(1)} + \kappa^2 \mathrm{H}_2'^{(2)} - E_2 \mathrm{P}_2 + \mathcal{O}(\kappa^3) \ . \tag{2.46}$$

As for the one-particle subspace, the diagonalization of the reduced Hamiltonian is most easily performed within the coordinate space representation. This means that $\mathcal{E}_2$ is identified with the space of normalizable wave functions on $\mathbb{Z}^d \times \mathbb{Z}^d$ which satisfy

$$\psi(x,y) = \psi(y,x) \ , \quad \psi(x,x) = 0 \ . \tag{2.47}$$

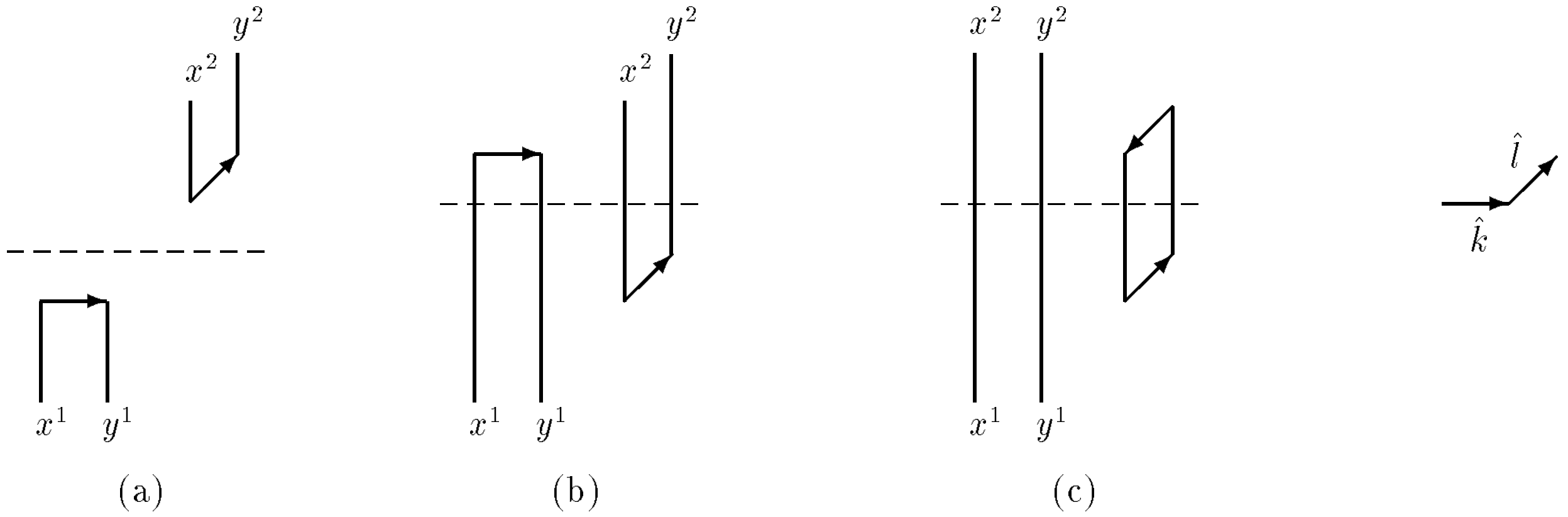

Figure 4: Next-to-leading order graphs with initial state $|x^1, y^1\rangle^0$ and final state $|x^2, y^2\rangle^0$.

We extend the definition (2.29) of the difference operators by a subscript that indicates which argument of the wave function is shifted

$$(\nabla_{x,k}\psi)(x,y) = \psi(x+\hat{k}, y) - \psi(x,y) \ , \quad (\nabla_{y,k}\psi)(x,y) = \psi(x, y+\hat{k}) - \psi(x,y) \ , \tag{2.48}$$

and correspondingly for $\nabla^*_{x,k}$ and $\nabla^*_{y,k}$. Moreover it is convenient to introduce

$$\nabla_{\eta,k,l} \stackrel{\text{def}}{=} \nabla_{\eta,k}\nabla_{\eta,l} + \nabla_{\eta,k} + \nabla_{\eta,l} \ , \quad \nabla^*_{\eta,k,l} \stackrel{\text{def}}{=} -\nabla^*_{\eta,k}\nabla^*_{\eta,l} + \nabla^*_{\eta,k} + \nabla^*_{\eta,l} \ , \tag{2.49}$$

where $\eta$ stands for $x$ or $y$, respectively. Finally, for any $\eta \in \mathbb{Z}^d$ we define the multiplication operator

$$(\mathrm{D}_\eta\psi)(x,y) = \delta(x-y-\eta)\psi(x,y) \ , \tag{2.50}$$

where $\delta(x)$ denotes the lattice $\delta$-function which simply is Kronecker's $\delta$ in case that the lattice spacing is equal to 1.

Collecting everything together the reduced two-particle Hamiltonian reads

$$\mathrm{H}_2 = \kappa(\mathrm{T}_2 + \mathrm{W}_2) = \kappa\left\{\mathrm{T}_2^{(1)} + \mathrm{W}_2^{(1)} + \kappa\left[\mathrm{T}_2^{(2)} + \mathrm{W}_2^{(2)}\right]\right\} + \mathcal{O}(\kappa^3) \ , \tag{2.51}$$

where

$$\begin{aligned}
\mathrm{T}_2^{(1)} &= \Delta_x + \Delta_y - 4d \ , \\
\mathrm{W}_2^{(1)} &= \mathrm{D}_0\lambda - \left(\mathrm{D}_0\mathrm{T}_2^{(1)} + \text{h.c.}\right) \ , \\
\mathrm{T}_2^{(2)} &= -\frac{1}{4}\left(\Delta_x^2 + \Delta_y^2\right) + d\left(\Delta_x + \Delta_y\right) - 2d(d-1) \ , \\
\mathrm{W}_2^{(2)} &= d\mathrm{D}_0 + (2d-1)\sum_{k=1}^{d}(\mathrm{D}_{\hat{k}} + \mathrm{D}_{-\hat{k}}) - \left\{\mathrm{D}_0\mathrm{T}_2^{(2)} - \frac{1}{4}\sum_{k,l=1}^{d}\left[\mathrm{D}_{\hat{k}}\left(\nabla_{y,k,l} - \nabla^*_{x,k,l}\right)\right.\right. \\
&\quad \left.\left. + \mathrm{D}_{-\hat{k}}\left(\nabla_{x,k,l} - \nabla^*_{y,k,l}\right) + \mathrm{D}_{\hat{k}}\left(\nabla_{y,k,-l} - \nabla^*_{x,k,-l}\right) + \mathrm{D}_{-\hat{k}}\left(\nabla_{x,k,-l} - \nabla^*_{y,k,-l}\right)\right] + \text{h.c.}\right\} \ .
\end{aligned}$$

The first part of $\mathrm{H}_2$ is just the sum of two effective one-particle Hamiltonians (2.32). Therefore $\mathrm{T}_2$ is identified with the kinectic term which determines the free propagation of two independent one-particle states. They interact through the potential $\mathrm{W}_2$ which depends on the relative coordinate of the two-particles $(x-y)$ and is rather short range.

However, we note that due to the restriction for coinciding arguments, $\mathcal{E}_2$ is *not* the appropriate two-particle space. In particular, it is not possible to represent a state that describes two independent one-particle wave functions moving freely. Instead, the natural Hilbert space for the evolution of a two-spin state is rather the symmetric tensor product of the one-particle subspace $\mathcal{E}_1$ with itself. This space is represented by the symmetric wave functions without any further restriction and will be denoted by $\mathcal{H}_2$. Since $\mathcal{E}_2$ is a closed subspace, $\mathcal{H}_2$ is the direct sum of $\mathcal{E}_2$ and its orthogonal complement $\mathcal{E}_2^\perp$, namely the space of symmetric wave functions which vanish except for coinciding arguments.

Now the discussion of scattering is much easier if the free and the interacting dynamics are defined on the same Hilbert space. Otherwise one would have to use the formalism of two Hilbert space scattering theory as described in ref. [18], for example. Though conceptually satisfactory this method bears some complications in practical calculations. It is therefore convenient to extend $\mathrm{H}_2'$ to a self-adjoint operator $\mathrm{H}_2$ on $\mathcal{H}_2$. Since $\mathrm{D}_0$ projects on $\mathcal{E}_2^\perp$ such an operator is given by

$$\mathrm{H}_2 = (\mathbb{1} - \mathrm{D}_0)\,\mathrm{H}_2'\,(\mathbb{1} - \mathrm{D}_0) + \kappa\lambda \mathrm{D}_0 \ , \tag{2.52}$$

where $\lambda$ is an arbitrary real constant. Since the free Hamiltonian $\mathrm{T}_2$ trivially extends to an operator on $\mathcal{H}_2$ we can rewrite (2.52) as

$$\mathrm{H}_2 = \kappa(\mathrm{T}_2 + \mathrm{V}_2) \ , \tag{2.53}$$

with a modified but still very short range potential

$$\mathrm{V}_2 = -\mathrm{D}_0\mathrm{T}_2 - \mathrm{T}_2\mathrm{D}_0 + \mathrm{D}_0\mathrm{T}_2\mathrm{D}_0 + (\mathbb{1} - \mathrm{D}_0)\mathrm{W}_2(\mathbb{1} - \mathrm{D}_0) + \lambda\mathrm{D}_0 \ . \tag{2.54}$$

The wave functions $\psi \in \mathcal{E}_2^\perp$ are rapidly decaying. Consequently, they do not belong to the absolutely continuous spectrum but represent bound states with energy $\kappa\lambda$. Therefore all physical scattering quantities must eventually be independent of $\lambda$.

Here we make a technical remark. Since the restriction to the symmetric wave functions complicates the diagonalization of the Hamiltonian (2.53), we will henceforth consider $\mathrm{H}_2$ as an operator on the entire space of normalizable wave functions $L^2(\mathbb{Z}^d \times \mathbb{Z}^d)$. This is possible because $\mathrm{H}_2$ is identical on $\mathcal{H}_2$ and $L^2(\mathbb{Z}^d \times \mathbb{Z}^d)$, and commutes with the unitary operator $\mathrm{U}_{12}$ representing the permutation $(\mathrm{U}_{12}\psi)(x_1, x_2) = \psi(x_2, x_1)$. So after we have succeeded in finding the eigenstates of the Hamiltonian (2.53) defined on $L^2(\mathbb{Z}^d \times \mathbb{Z}^d)$ the true two-particle sector of the Ising model is restored by projecting on the subspace of symmetric wave functions.

Since the two-body Hamiltonian (2.51) commutes with the translation operator $(\mathrm{U}_v\psi)(x, y) = \psi(x + v, y + v)$ the center of mass motion can be separated. Let $p_x$ and $p_y$ denote the momenta corresponding to $x$ and $y$, respectively. Introducing

$$r = x - y \ , \quad q = \frac{1}{2}(p_x - p_y) \quad \text{and} \quad R = x + y \ , \quad P = \frac{1}{2}(p_x + p_y) \ , \tag{2.55}$$

the simultaneous (improper) eigenstates of $\mathrm{H}_2$ and $\mathrm{U}_v$ have the general form

$$\psi(x, y) = \mathrm{e}^{\mathrm{i}PR}\widehat{\psi}(r) \ , \tag{2.56}$$

where $\widehat{\psi} \in L^2(\mathbb{Z}^d)$. Now if we introduce

$$\mathrm{H}(P) \stackrel{\text{def}}{=} \mathrm{e}^{\mathrm{i}PR}\mathrm{H}_2\mathrm{e}^{-\mathrm{i}PR} \ , \tag{2.57}$$

the action of $\mathrm{H}_2$ is expressed as

$$(\mathrm{H}_2\psi)(x, y) = \mathrm{e}^{\mathrm{i}PR}\left(\mathrm{H}(P)\widehat{\psi}\right)(r) \ . \tag{2.58}$$

So the eigenvalue problem of the two-particle Hamiltonian $\mathrm{H}_2$ is reduced to the diagonalization of the effective one-body Hamiltonian $\mathrm{H}(P)$. The corresponding Hilbert space $\mathcal{H}(P)$ is a translationally invariant subspace of $L^2(\mathbb{Z}^d \times \mathbb{Z}^d)$. It is identified with the space of normalizable wave functions in the relative coordinate $r$. For simplicity we will restrict our attention to the center of mass system, i.e. in the following we assume that the total momentum $P$ of the two particles is zero.

The explicit calculation of (2.57) is straightforward. We state the final result

$$\mathrm{H} = \kappa(\mathrm{T} + \mathrm{V}) \tag{2.59}$$

where the free Hamiltonian reads

$$\mathrm{T} = 2\{\Delta_r - 2d\} - \frac{\kappa}{2}\{\Delta_r^2 - 4d\Delta_r + 4d(d-1)\} + \mathcal{O}(\kappa^2)\ , \tag{2.60}$$

and the potential is given by

$$\mathrm{V} = \mathrm{V}^{(1)} + \kappa\mathrm{V}^{(2)} + \mathcal{O}(\kappa^2)\ , \tag{2.61}$$

$$\mathrm{V}^{(1)} = -\mathrm{D}_0\mathrm{T} - \mathrm{T}\mathrm{D}_0 + \mathrm{D}_0\mathrm{T}\mathrm{D}_0 + \lambda\mathrm{D}_0\ , \tag{2.62}$$

$$\begin{aligned}\mathrm{V}^{(2)} = {} & (2d-1)\sum_k^d \{\mathrm{D}_{\hat{k}} + \mathrm{D}_{-\hat{k}}\} \\ & -\frac{1}{2}\sum_{k,l=1}^d \left\{\mathrm{D}_{\hat{k}}\nabla^*_{r,k,l} - \mathrm{D}_{-\hat{k}}\nabla_{r,k,l} + \mathrm{D}_{\hat{k}}\nabla^*_{r,k,-l} - \mathrm{D}_{-\hat{k}}\nabla_{r,k,-l} + \mathrm{h.c.}\right\}\ .\end{aligned} \tag{2.63}$$

# 3 The scattering formalism

The two-particle dynamics of the $(d+1)$-dimensional Ising model is determined by the effective Hamiltonian (2.59). To obtain the desired strong coupling expansion for the elastic scattering states and the transition matrix we proceed following the rules of ordinary quantum mechanics.

## 3.1 Brief review of quantum mechanical scattering theory

The starting point of scattering theory is the physical assumption that the constituents in the initial and final state are so far away from each other that the interaction between them is negligible. In our case this means that the scattering states of the effective one-body system (2.59) are those vectors in the Hilbert space which are 'asymptotically' free in the distant past and future. The mathematically precise meaning of this statement is that for any $f$ in the absolutely continuous subspace $\mathcal{H}_{\mathrm{ac}}(\mathrm{T})$ of the free Hamiltonian T there is a vector $f^{\mathrm{in}} \in L^2(\mathbb{Z}^d)$ such that

$$\lim_{t\to-\infty} \| \,\mathrm{e}^{-\mathrm{i}\mathrm{T}t} f - \mathrm{e}^{-\mathrm{i}\mathrm{H}t} f^{\mathrm{in}}\, \| = 0\ ,$$

and similiarly for pairs of states $\{f, f^{\mathrm{out}}\}$ that approach each other as $t \to +\infty$. Let $\mathrm{P}_{\mathrm{ac}}(\mathrm{T})$ denote the projection operator on $\mathcal{H}_{\mathrm{ac}}(\mathrm{T})$. Then the basic existence question of scattering is reduced to the problem of proving that the Møller operators

$$\Omega_{\mathrm{in}}(\mathrm{H},\mathrm{T}) \stackrel{\mathrm{def}}{=} \mathrm{s}-\lim_{t\to-\infty} \mathrm{e}^{\mathrm{i}\mathrm{H}t}\mathrm{e}^{-\mathrm{i}\mathrm{T}t}\mathrm{P}_{\mathrm{ac}}(\mathrm{T})\ , \tag{3.1}$$

$$\Omega_{\mathrm{out}}(\mathrm{H},\mathrm{T}) \stackrel{\mathrm{def}}{=} \mathrm{s}-\lim_{t\to+\infty} \mathrm{e}^{\mathrm{i}\mathrm{H}t}\mathrm{e}^{-\mathrm{i}\mathrm{T}t}\mathrm{P}_{\mathrm{ac}}(\mathrm{T}) \tag{3.2}$$

exist. In case they do, in- and out-going scattering states are defined as

$$f^{\mathrm{in}} = \Omega_{\mathrm{in}} f\ ,\quad f^{\mathrm{out}} = \Omega_{\mathrm{out}} f\ . \tag{3.3}$$

Furthermore we expect that the ranges $\mathcal{H}_{\mathrm{in/out}}$ of the in- and out-going Møller operators are equal to the absolutely continuous subspace $\mathcal{H}_{\mathrm{ac}}(\mathrm{H})$ because any interacting state which looked like a free state in the distant past should again look like a free state for large positive times. This property is called the completeness of the Møller operators.

To prove that $\Omega_{\rm in}$ and $\Omega_{\rm out}$ exist and are complete we take advantage of the special structure of the potential. From the derivation of the reduced two-particle Hamiltonian by means of Bloch's perturbation theory it should be clear that to *all orders* of $\kappa$ the effective potential V will be a sum of operators each of which is the product of a constant or a difference operator and a local multiplication operator $\mathrm{D}_\eta$. Now any difference operator on $L^2(\mathbb{Z}^d)$ is bounded. Furthermore, $\mathrm{D}_\eta$ is trace class because it is a positive operator and satisfies

$$\mathrm{Tr}(\mathrm{D}_\eta) = \sum_{\nu=1}^{\infty}(\varphi_\nu, \mathrm{D}_\eta \varphi_\nu) = \sum_{\nu=1}^{\infty}\sum_r \varphi_\nu(r)^*(\mathrm{D}_\eta\varphi_\nu)(r) = \sum_{\nu=1}^{\infty}\varphi_\nu(\eta)^*\varphi_\nu(\eta) = 1$$

for any orthonormal basis $\{\varphi_\nu\}_{\nu=1}^{\infty}$ of $L^2(\mathbb{Z}^d)$. The product of a bounded operator and a trace class operator is trace class and the family of trace class operators is a vector space (cf. theorem VI.19 of ref. [18]). It follows that the potential V is trace class, and the prerequisites of the Kato-Rosenblum theorem are fulfilled. This guarantees the existence and completeness of $\Omega_{\rm in/out}(\mathrm{H},\mathrm{T})$ (cf. theorem XI.8 of ref. [18]).

The fundamental operator of scattering is

$$\mathsf{S} \stackrel{\rm def}{=} \Omega_{\rm out}\,(\Omega_{\rm in})^\dagger \ : \ \mathcal{H}_{\rm in} \to \mathcal{H}_{\rm out}\ . \tag{3.4}$$

It maps in-going states onto out-going states with the same probability distribution of momentum. Hence the scattering operator correlates the past and the future asymptotics and determines the physics of the scattering process. The scattering operator commutes with the Hamiltonian H and is unitary on $\mathcal{H}_{\rm ac}(\mathrm{H})$.

The free Hamiltonian T has the purely continuous spectrum

$$\kappa\epsilon_2(q) = \epsilon_1(p_x) + \epsilon_1(p_y) \tag{3.5}$$

where

$$\epsilon_2(q) = 2\left\{\widehat{q}^2 - 2d\right\} - \frac{\kappa}{2}\left\{(\widehat{q}^2)^2 - 4d\widehat{q}^2 + 4d(d-1)\right\} + \mathcal{O}(\kappa^2)\ . \tag{3.6}$$

From [15] we know that the series representation for T is convergent if the inverse coupling is sufficently small. It follows that the two-particle energy is a smooth function of the relative momentum $q$. The plane waves

$$\Phi(q;r) = \frac{\mathrm{e}^{\mathrm{i}qr}}{(2\pi)^{\frac{d}{2}}}\ , \quad -\pi \le q \le \pi \tag{3.7}$$

are a complete set of (improper) eigenfunctions normalized such that $(\Phi(q'),\Phi(q)) = \delta(q'-q)$. In the sense of distributions (improper) in- and out-going scattering states can be defined as

$$\Phi^{\rm in}(q) \stackrel{\rm def}{=} \Omega_{\rm in}\Phi(q)\ , \quad \Phi^{\rm out}(q) \stackrel{\rm def}{=} \Omega_{\rm out}\Phi(q)\ . \tag{3.8}$$

They are (improper) eigenstates of the Hamiltonian H with energy $\kappa\epsilon_2(q)$. They span the absolutely continuous subspace $\mathcal{H}_{\rm ac}(\mathrm{H})$ and satisfy

$$(\Phi^{\rm in}(q'),\Phi^{\rm in}(q)) = (\Phi^{\rm out}(q'),\Phi^{\rm out}(q)) = (\Phi(q'),\Phi(q))\ .$$

Starting from (3.8) one deduces an integral equation for the scattering solutions, the famous Lippmann-Schwinger equations

$$\Phi^{\rm in/out}(q) = \Phi(q) - \lim_{\rho\searrow 0}\frac{1}{\mathrm{T}-\epsilon_2(q)\mp\mathrm{i}\rho}\,\mathrm{V}\Phi^{\rm in/out}(q)\ . \tag{3.9}$$

With the solutions of these equations at hand normalizable scattering states are obtained by smearing with a smooth periodic function in the relative momentum $q$

$$f^{\rm in} = \int_{\mathcal{B}} \mathrm{d}^d q \tilde{f}(q)\Phi^{\rm in}(q)\ , \quad f^{\rm out} = \int_{\mathcal{B}} \mathrm{d}^d q \tilde{f}(q)\Phi^{\rm out}(q)\ .$$

This leads to an integral representation of the scattering operator

$$(f^{\text{in}}, \mathsf{S}f^{\text{in}}) = \int_{\mathcal{B}} \mathrm{d}^d q' \int_{\mathcal{B}} \mathrm{d}^d q \tilde{f}(q')^* \tilde{f}(q) \, (\Phi^{\text{in}}(q'), \mathsf{S}\Phi^{\text{in}}(q)) \ ,$$

and the physics of the scattering process is determined by its kernel, the $\mathsf{S}$-matrix

$$\mathsf{S}(q', q) = (\Phi^{\text{in}}(q'), \mathsf{S}\Phi^{\text{in}}(q)) = (\Phi^{\text{in}}(q'), \Phi^{\text{out}}(q)) \ . \tag{3.10}$$

By means of the Lippman-Schwinger equations one can establish an explicit formula

$$\mathsf{S}(q', q) = (\Phi(q'), \Phi(q)) - 2\pi \mathrm{i}\kappa \, \delta[E_2(q') - E_2(q)] \mathbb{T}(q', q) \ , \tag{3.11}$$

where $E_2(q)$ is the full two-particle energy

$$E_2(q) = 4 + \kappa \epsilon_2(q) \ , \tag{3.12}$$

and the on-shell transition matrix is defined as

$$\mathbb{T}(q', q) = (\Phi(q'), \mathrm{V}\Phi^{\text{in}}(q)) \ . \tag{3.13}$$

Finally we have to remember that only the *symmetric* wave functions belong to the true two-particle sector of our model. Since the permutation operator $\mathrm{U}_{12}$ commutes with both T and H, it also commutes with the Møller operators (3.1), (3.2) and hence with the scattering operator (3.4). This means that in (3.11) and (3.13) we simply have to substitute the wave functions by their symmetric counterparts

$$\Phi_{\text{s}}(q; r) \stackrel{\text{def}}{=} \frac{1}{2}\{\Phi(q; r) + \Phi(q; -r)\} = \frac{\cos(qr)}{(2\pi)^{\frac{d}{2}}} \ , \tag{3.14}$$

$$\Phi_{\text{s}}^{\text{in}}(q, r) = \frac{1}{2}\{\Phi^{\text{in}}(q; r) + \Phi^{\text{in}}(q; -r)\} \ , \tag{3.15}$$

which satisfy

$$(\Phi_{\text{s}}(q'), \Phi_{\text{s}}(q)) = (\Phi_{\text{s}}^{\text{in}}(q'), \Phi_{\text{s}}^{\text{in}}(q)) = \frac{1}{2}\{\delta(q' - q) + \delta(q' + q)\} \ . \tag{3.16}$$

## 3.2 Generalized Born series

It is convenient to introduce the Green's operator

$$\mathrm{G}(q) = \lim_{\rho \searrow 0} \frac{1}{\epsilon_2(q) - \mathrm{T} + \mathrm{i}\rho} \ . \tag{3.17}$$

Since the free Hamiltonian T commutes with the unitary operator $\mathrm{U}_R$ representing the hypercubic transformations

$$(\mathrm{U}_R \psi)(r) = \psi(Rr) \ , \quad R \in \mathrm{O}(d, \mathbb{Z}) \ , \tag{3.18}$$

so does $\mathrm{G}(q)$. From now on we follow the usual convention that $\rho > 0$ is understood as the limit $\rho \to 0$. In coordinate space (3.17) is referred to as the Green's function

$$G(q; r, r') = {}^0\langle r | \mathrm{G}(q) | r' \rangle^0 \tag{3.19}$$

For fixed $q$ it is a well defined function on $(\mathbb{Z}^d \times \mathbb{Z}^d)$ which is invariant under identical hypercubic transformations in both arguments and has an integral representation as

$$G(q; r, r') = \int_{\mathcal{B}} \frac{\mathrm{d}^d p}{(2\pi)^d} \frac{\mathrm{e}^{\mathrm{i}p(r - r')}}{\epsilon_2(q) - \epsilon_2(p) + \mathrm{i}\rho} \tag{3.20}$$

Since the energy $\epsilon_2(q)$ is a smooth hypersurface in momentum space the Green's function is a smooth function of the relative momentum $q$, at least as long as the energy hypersurface does not intersect with the boundary of the Brillouin zone. For notational convenience we write $G(q;r)$ instead of $G(q;r,0)$.

Using (3.17) the Lippmann-Schwinger equation for the in-going scattering states reads

$$\Phi^{\rm in}(q) = \Phi(q) + {\rm G}(q)\,{\rm V}\,\Phi^{\rm in}(q)\ . \qquad (3.21)$$

The potential is a convergent power series in the inverse coupling where the range of $V$ increases with increasing order in $\kappa$. We are heading now for a solution of the above integral equation[1]. The first step is to decompose (3.21) as follows

$$\Phi^{\rm in}(q) = \Phi(q) + {\rm G}(q)\,{\rm V}^{(1)}\,\Phi^{\rm in}(q) + {\rm G}(q)\,\left[{\rm V} - {\rm V}^{(1)}\right]\,\Phi^{\rm in}(q)\ . \qquad (3.22)$$

Since for the leading order potential $V^{(1)} = \mathcal{O}(1)$ the second term is of the same order as the free solution $\Phi(q)$, whereas the third contribution is proportional to $\kappa$. In the next subsection we show that

$$\Phi_1^{\rm in}(q) = \Phi(q) + {\rm G}(q)\,{\rm V}^{(1)}\,\Phi_1^{\rm in}(q) \qquad (3.23)$$

can be solved exactly thanks to the locality of $V^{(1)}$. Substituting this result, eq. (3.22) becomes

$$\Phi^{\rm in}(q) = \Phi_1^{\rm in}(q) + \left[\mathbb{1} - {\rm G}(q){\rm V}^{(1)}\right]^{-1}{\rm G}(q)\left[{\rm V} - {\rm V}^{(1)}\right]\Phi^{\rm in}(q)\ . \qquad (3.24)$$

Thereby we have derived a modified integral equation for the scattering solution. To solve eq. (3.24) we make a Born series ansatz

$$\Phi^{\rm in}(q) = \sum_{\nu=0}^{\infty}\left\{\left[\mathbb{1} - {\rm G}(q){\rm V}^{(1)}\right]^{-1}{\rm G}(q)\left[{\rm V} - {\rm V}^{(1)}\right]\right\}^{\nu}\Phi_1^{\rm in}(q) \qquad (3.25)$$

which represents a power series in the inverse coupling. For a discussion of convergence we consider (3.25) in the coordinate space representation. Since all contributions to the potential $V$ have finite support there are no infinite sums but products of Green's functions evaluated at definite coordinate points. To any order $\mathcal{O}(\kappa^n)$ these products consists of $n$ factors at most. In addition the Green's function is bounded. So there is no doubt that the Born series converges if the inverse coupling is sufficiently small. Consequently, (3.25) is indeed a solution of the Lippmann-Schwinger equation (3.21) and a valid representation of the Ising model's scattering states.

## 3.3 Leading order

We consider the leading order Lippmann-Schwinger equation (3.23) in the coordinate space representation

$$\Phi_1^{\rm in}(q;r) = \Phi(q;r) + \sum_{r} {\rm G}(q;r,r'){\rm V}^{(1)}\Phi_1^{\rm in}(q;r')\ . \qquad (3.26)$$

To evaluate the r.h.s. we substitute the explicit form of the leading order potential (2.62). By means of the integral representation (3.20) one verifies the difference equation

$$\{-{\rm T} + \epsilon_2(q)\}\,G(q;r) = \delta(r)\ . \qquad (3.27)$$

for the Green's function. It follows that

$$\sum_{r'} G(q;r,r')\,{\rm T}\,{\rm D}_0\,\Phi_1^{\rm in}(q;r') = z_1^{(1)}\{\epsilon_2(q)G(q;r) - \delta(r)\}\ . \qquad (3.28)$$

[1] From what follows it will be clear that we can derive a solution for the out-going states analogously.

where we have abbreviated $z_1^{(1)} = \Phi_1^{\rm in}(q;0)$ Furthermore, inserting a complete set of plane waves we obtain

$$\mathrm{D}_0\,\mathrm{T}\,\mathrm{D}_0\,\Phi_1^{\rm in}(q;r) = \underbrace{\int_{\mathcal{B}} \frac{\mathrm{d}^d p}{(2\pi)^d}\epsilon_2(p)}_{\omega}\; z_1^{(1)}\,\delta(r)\;. \tag{3.29}$$

Finally the action of $\mathrm{D}_0\,\mathrm{T}$ on the wave function $\Phi_1^{\rm in}(q;r)$ leads to a a sum of values of $\Phi_1^{\rm in}(q;r)$ at definite coordinate points $r$ which we abbreviate by a complex number $z_2^{(1)}$

$$\mathrm{D}_0\,\mathrm{T}\,\Phi_1^{\rm in}(q;r) = z_2^{(1)}\,\delta(r)\;. \tag{3.30}$$

Substituting (3.28)–(3.30), eq. (3.26) becomes

$$\Phi_1^{\rm in}(q;r) = \Phi(q;r) + z_1^{(1)}\,\delta(r) - \left\{ z_2^{(1)} + [\epsilon_2(q) - \omega - \lambda]\,z_1^{(1)} \right\} G(q;r)\;. \tag{3.31}$$

So the coordinate dependence of the leading order scattering states is determined by the free solution $\Phi(q;r)$ and the Green's function $G(q;r)$. To determine the values of the unknown numbers $z_1^{(1)}$ and $z_2^{(1)}$ we first evaluate (3.31) at $r=0$

$$\{\epsilon_2(q) - \omega - \lambda\}G(q;0)\,z_1^{(1)} + G(q;0)\,z_2^{(1)} = \Phi(q;0)\;, \tag{3.32}$$

and then act on (3.31) by $\mathrm{D}_0\mathrm{T}$

$$\{\lambda - \epsilon_2(q) + [\epsilon_2(q) - \omega - \lambda]\epsilon_2(q)G(q;0)\}\,z_1^{(1)} + \epsilon_2(q)G(q;0)\,z_2^{(1)} = \epsilon_2(q)\Phi(q;0)\;. \tag{3.33}$$

Combining (3.32) and (3.33) one obtains a simple matrix equation for $z_1^{(1)}$ and $z_2^{(1)}$ which can be solved easily

$$z_1^{(1)} = 0\;, \quad z_2^{(1)} = \left[(2\pi)^{d/2}G(q;0)\right]^{-1}\;. \tag{3.34}$$

In particular we find that the scattering solutions vanishes at $r=0$. We substitute (3.34) into (3.31) and symmetrize the result. The leading order scattering solution reads

$$\Phi_{\mathrm{s},1}^{\rm in}(q;r) = \Phi_{\mathrm{s}}(q;r) - \frac{G(q;r)}{(2\pi)^{\frac{d}{2}}G(q;0)}\;. \tag{3.35}$$

With this solution at hand it is also straightforward to calculate the leading order on-shell transition matrix (cf. eq (3.13)

$$\mathbb{T}_1(q',q) = \left(\Phi_{\mathrm{s}}(q'), \mathrm{V}^{(1)}\Phi_{\mathrm{s},1}^{\rm in}(q)\right) = -\left[(2\pi)^d G(q;0)\right]^{-1}\;. \tag{3.36}$$

The reader should not be frightened at the apparent violation of time reflection symmetry $q' \leftrightarrow q$. From (3.20) it follows that $G(q';r) = G(q,r)$ if $\epsilon_2(q') = \epsilon_2(q)$.

## 3.4 Next-to-leading order

The next-to-leading order contribution to the Born series (3.25) reads

$$\Phi_2^{\rm in}(q) = \left[\mathbb{1} - \mathrm{G}(q)\mathrm{V}^{(1)}\right]^{-1}\mathrm{G}(q)\left[\mathrm{V} - \mathrm{V}^{(1)}\right]\Phi_1^{\rm in}(q)\;. \tag{3.37}$$

In coordinate space we obtain the integral equation

$$\Phi_2^{\rm in}(q;r) = \kappa A(q;r) + \sum_{r'} \mathrm{G}(q;r,r')\mathrm{V}^{(1)}\Phi_2^{\rm in}(q;r') \tag{3.38}$$

where we have abbreviated

$$A(q;r) = \sum_{r'} G(q;r,r')\mathrm{V}^{(2)}\Phi_1^{\rm in}(q;r')\;, \tag{3.39}$$

and the next-to-leading order potential $V^{(2)}$ is given by (2.63). Except for the inhomogeneity, eq. (3.38) has exactly the same structure as the leading order Lippmann-Schwinger equation (3.26), hence

$$\Phi_2^{\rm in}(q;r) = \kappa A(q;r) + z_1^{(2)}\,\delta(r) - \left\{ z_2^{(2)} + [\epsilon_2(q) - \omega - \lambda]\, z_1^{(2)} \right\} G(q;r) \;, \tag{3.40}$$

where

$$z_1^{(2)} = \Phi_2^{\rm in}(q;0) \;, \quad {\rm D}_0\,{\rm T}\,\Phi_2^{\rm in}(q;r) = z_2^{(2)}\,\delta(r) \;. \tag{3.41}$$

Substituting the leading order solution (3.35) and the potential (2.63) it is straightforward to calculate

$$\begin{aligned} A(q,r) \;&=\; -\frac{1}{(2\pi)^{d/2}} \left\{ \hat{q}^2 - 2d + \frac{(2d-1)G(q;\hat{1})}{G(q,0)} \right\} \sum_k \left\{ G(q;r-\hat{k}) + G(q;r+\hat{k}) \right\} \\ &\quad -\frac{1}{(2\pi)^{d/2}} \sum_k \cos(q_k) \left\{ G(q;r-\hat{k}) + G(q;r+\hat{k}) \right\} \\ &\quad -\frac{\rm i}{(2\pi)^{d/2}} \sum_k \sin(q_k) \left\{ G(q;r-\hat{k}) - G(q;r+\hat{k}) \right\} \;. \end{aligned} \tag{3.42}$$

Here and below we make use of the identity $G(q;\pm\hat{k}) = G(q;\hat{1})$ which is a consequence of the hypercubic symmetry of the Green's function.

Since the coordinate dependence of the inhomogeneity $A(q;r)$ is again determined by the Green's function, eq. (3.40) can be solved exactly in the same way as (3.31). Using (3.27) one finds

$${\rm D}_0{\rm T}A(q;r) = \epsilon_2(q) A(q;0)\,\delta(r) \;. \tag{3.43}$$

So the resulting equations that fix the unknown variables $z_1^{(2)}$ and $z_2^{(2)}$ are the same as (3.32) and (3.33) if we only replace $\Phi(q;0)$ by $\kappa A(q;0)$. It follows

$$z_1^{(2)} = 0 \;, \quad z_2^{(2)} = \kappa \frac{A(q;0)}{G(q;0)} \;. \tag{3.44}$$

Substituting these values into (3.40) and symmetrizing the result we end up with

$$\Phi_{2,{\rm s}}^{\rm in}(q;r) = \kappa \left\{ A_{\rm s}(q;r) - \frac{A_{\rm s}(q;0)}{G(q;0)}\, G(q;r) \right\} \;, \tag{3.45}$$

where

$$\begin{aligned} A_{\rm s}(q,r) \;&=\; -\frac{1}{(2\pi)^{d/2}} \left\{ \hat{q}^2 - 2d + \frac{(2d-1)G(q;\hat{1})}{G(q,0)} \right\} \sum_k \left\{ G(q;r-\hat{k}) + G(q;r+\hat{k}) \right\} \\ &\quad -\frac{1}{(2\pi)^{d/2}} \sum_k \cos(q_k) \left\{ G(q;r-\hat{k}) + G(q;r+\hat{k}) \right\} \;, \end{aligned} \tag{3.46}$$

$$A_{\rm s}(q,0) \;=\; -\frac{(2d-1)G(q;\hat{1})}{(2\pi)^{d/2}} \left\{ \hat{q}^2 - 2d + 2d\frac{G(q;\hat{1})}{G(q;0)} \right\} \;. \tag{3.47}$$

The next-to-leading order contribution to the transition matrix consists of two parts

$$\mathbb{T}_2(q',q) = \left( \Phi_{\rm s}(q'), {\rm V}^{(1)}\Phi_{{\rm s},2}^{\rm in}(q) \right) + \kappa \left( \Phi_{\rm s}(q'), {\rm V}^{(2)}\Phi_{{\rm s},1}^{\rm in}(q) \right) \;. \tag{3.48}$$

Since $\Phi_{2,{\rm s}}^{\rm in}(q;0) = 0$ the first term reduces to

$$\mathbb{T}_2^{(1)}(q',q) = -\left( \Phi_{\rm s}(q'), {\rm D}_0{\rm T}\Phi_{{\rm s},2}^{\rm in}(q) \right) \;. \tag{3.49}$$

which is calculated trivially using the definition of $z_2^{(2)}$ (3.41) and the result (3.44). To determine the second contribution we simply have to insert the result for $\mathrm{V}^{(2)}\Phi^{\mathrm{in}}_{\mathrm{s},1}(q;r)$ obtained previously (cf. eqs. (3.39) and (3.42)). Combining everything together the final result reads

$$\begin{aligned}\mathbb{T}_2(q',q) &= -\frac{\kappa}{(2\pi)^d}\Bigg\{2\sum_k \cos(q_k)\cos(q'_k) - (\widehat{q}'^2 - 2d)(\widehat{q}^2 - 2d) - \frac{(2d-1)G(q;\hat{1})}{G(q;0)} \\ &\qquad\times\left[\widehat{q}^2 + \widehat{q}^2 - 4d + 2d\frac{G(q;\hat{1})}{G(q;0)}\right]\Bigg\} \ . \end{aligned} \tag{3.50}$$

The approximation for the scattering quantities presented here relies on the Born series representation of the scattering solution (3.25). It is strongly motivated by the physical structure of the effective one-body quantum system derived by means of Bloch's perturbation theory. Alternatively one could think of a polynomial approximation in the inverse coupling which would require to expand the Green's function (3.20) in powers of $\kappa$. This is perfectly alright here but demands some care if $G(q;r)$ is more complicated. It might even happen that the Green's function is singular at $\kappa = 0$. Anyhow, a polynomial approximation for the scattering quantities is less natural and will be worse in general.

# 4 The $(1+1)$-dimensional model

In the preceding section we have derived the general formulae for the leading and next-to-leading order contributions to the scattering solution and the transition matrix. To get a feeling of the results obtained we specialize to the simplest case of $d = 1$ spatial dimension. For this model exact solutions exist [19] and it is therefore a good testing ground for perturbative methods.

The next-to-leading order approximation for the two-particle energy (3.6) can be written as

$$E_2(q) = 4 - 4\kappa\cos q + 2\kappa^2\sin^2 q \ . \tag{4.1}$$

As long as the inverse coupling is smaller than the critical value $\kappa_c = 1$ the denominator of the Green's function (3.20) factorizes

$$G(q;r) = \frac{1}{2\kappa}\int_{-\pi}^{\pi}\frac{\mathrm{d}p}{2\pi}\frac{\mathrm{e}^{\mathrm{i}pr}}{(\cos p - \cos q + \mathrm{i}\rho)(\cos p + \cos q + 2/\kappa)} \ . \tag{4.2}$$

For fixed $q$ and $r$ the integrand extends to an analytic function of $p$ in the domain $\mathbb{C}\setminus\{p_1,p_2\}$ where

$$p_1 = |q| - \frac{\mathrm{i}\rho}{\sin|q|}\ (\mathrm{mod}\,2\pi)\ , \quad p_2 = \pi + \mathrm{i}\,\mathrm{arcosh}(\cos q + 2/\kappa)\ (\mathrm{mod}\,2\pi)\ .$$

By means of the fundamental theorem in the calculus of residues one obtains

$$G(q;r) = \frac{-\mathrm{i}}{4(1+\kappa\cos q)}\left\{\frac{\mathrm{e}^{\mathrm{i}|q||r|}}{\sin|q|} - \mathrm{i}\kappa\frac{(-1)^r\mathrm{e}^{-\mu|r|}}{\bar{\mu}}\right\} \ . \tag{4.3}$$

where we have abbreviated

$$\mu = \mathrm{arcosh}(\cos q + 2/\kappa)\ , \quad \bar{\mu} = \kappa\sinh(\mu) = \sqrt{4(1+\kappa\cos q) - \kappa^2\sin^2 q} \ . \tag{4.4}$$

With this result at hand it is now an easy task to determine the next-to-leading order approximation of the scattering solution. Using (3.35) and (3.45) we find

$$\Phi^{\mathrm{in}}_{\mathrm{s}}(q;r) = \frac{1}{\sqrt{2\pi}}\left\{\cos(qr) + c_1\mathrm{e}^{\mathrm{i}|q||r|} + c_2(-1)^r\mathrm{e}^{-\mu|r|}\right\} + \mathcal{O}(\kappa^2) \ , \tag{4.5}$$

where the coefficients $c_1$ and $c_2$ are given by

$$\begin{aligned} c_1 &= \frac{-1}{\bar{\mu} - \mathrm{i}\kappa \sin|q|} \left\{ \bar{\mu} - \frac{2\mathrm{i}\kappa \sin|q|(1+\kappa\cos q)}{\bar{\mu} - \mathrm{i}\kappa \sin|q|} \right\} , \\ c_2 &= \frac{\mathrm{i}\kappa \sin|q|}{\bar{\mu} - \mathrm{i}\kappa \sin|q|} \left\{ 1 + \frac{2\mathrm{i}\kappa \sin|q|(1+\kappa\cos q)}{\bar{\mu} - \mathrm{i}\kappa \sin|q|} \right\} . \end{aligned} \tag{4.6}$$

In order to calculate the phase shifts we have to diagonalize the $\mathsf{S}$-matrix (3.11). Since the scattering operator commutes with the two-particle Hamiltonian it is favourable to introduce energy eigenstates $|E_2\rangle$ and $|E_2 \text{ in}\rangle$ which are normalized such that

$$\langle E_2' \mid E_2\rangle = \langle E_2' \text{ in} \mid E_2 \text{ in}\rangle = \delta(E_2' - E_2) . \tag{4.7}$$

The two-particle energy (4.1) is a even one-dimensional function of the relative momentum $q$. Consequently it is $E_2' = E_2$ if and only if $q' = \pm q$, hence

$$\delta(E_2' - E_2) = \left| \frac{\mathrm{d}E_2}{\mathrm{d}q} \right|^{-1} \{\delta(q'-q) + \delta(q'+q)\} . \tag{4.8}$$

It follows that the desired eigenstates are obtained if we define

$$|E_2\rangle = \left| \frac{1}{2}\frac{\mathrm{d}E_2}{\mathrm{d}q} \right|^{-1/2} \Phi_{\mathrm{s}}(q) , \quad |E_2 \text{ in}\rangle = \left| \frac{1}{2}\frac{\mathrm{d}E_2}{\mathrm{d}q} \right|^{-1/2} \Phi_{\mathrm{s}}^{\mathrm{in}}(q) . \tag{4.9}$$

With respect to these new bases of free and scattering states the scattering matrix takes the form

$$\mathsf{S}(E_2', E_2) = \left\{ 1 - 4\pi\mathrm{i}\kappa \left| \frac{\mathrm{d}E_2}{\mathrm{d}q} \right|^{-1} \mathbb{T}(E_2) \right\} \delta(E_2' - E_2) . \tag{4.10}$$

Using the general results (3.36) and (3.50) for the transition matrix we obtain the next-to-leading order approximation

$$\mathsf{S}(E_2', E_2) = \left\{ -\frac{4(1+\kappa\cos q)(1-\mathrm{i}\kappa\sin|q|)}{(\bar{\mu} - \mathrm{i}\kappa\sin|q|)^2} \right\} \delta(E_2' - E_2) + \mathcal{O}(\kappa^2) . \tag{4.11}$$

Due to the unitarity of the scattering matrix its eigenvalue has absolute value 1. Therefore it can be parametrized as $\exp(2\mathrm{i}\delta)$ where the angle $\delta$ is referred to as the phase shift or the scattering phase, respectively. For any relative momentum $q$ and any inverse coupling $\kappa < \kappa_{\mathrm{c}}$ eq. (4.11) allows an approximate calculation of the phase shift in the $(1+1)$-dimensional Ising model. Within errors we find the constant value

$$\delta = \frac{\pi}{2} . \tag{4.12}$$

In this simple model it is worth while to consider the polynomial approximation as well. The result is

$$\Phi_{\mathrm{s}}^{\mathrm{in}}(q; r) = \frac{-\mathrm{i}}{\sqrt{2\pi}} \sin(|q||r|) + \mathcal{O}(\kappa^2) \tag{4.13}$$

for the scattering solution and

$$\delta = \frac{\pi}{2} + \mathcal{O}(\kappa^2) . \tag{4.14}$$

for the phase shift. In fact (4.13) and (4.14) hold not only in the center of mass system but for any total momentum $P$ of the scattering particles. An interesting observation is of course that there is no $\mathcal{O}(\kappa)$ contribution. This result may be compared to the strong coupling series for the one-particle mass. From our general expression (2.35) it follows that for $d = 1$ there is no second order contribution either. Moreover, higher order calculations show that the series actually truncates after the first term and one finds that the leading order contribution represents the *exact* result for the one-particle mass [21]

$$m = 2(1 - \kappa) \ . \tag{4.15}$$

It may be that this is also the case for the scattering matrix meaning that $-1$ is indeed the *exact* result in the symmetric phase $\kappa < \kappa_c = 1$. The assumption is corroborated by recent studies of Lang et al. [7]. These authors determined the phase shift to be equal to $\pi/2$ by means of Lüscher's finite size technique approach [1] alluded to in the introduction. Anyhow, in ref. [22] it was shown that at the phase transition point $\kappa_c = 1$ the scattering matrix is $-1$.

We end our investigations of the $(1+1)$-dimensional model with a discussion of the physical properties of the derived scattering states. The (improper) eigenstates of the hermitian reduced two-particle Hamiltonian read

$$|p_x, p_y \text{ in}\rangle^0 = -\frac{\mathrm{i}}{2\pi} \sum_{x,y} \mathrm{e}^{\mathrm{i}P(x+y)} \sin(|q||x-y|)|x,y\rangle^0 + \mathcal{O}(\kappa^2) \ . \tag{4.16}$$

They are labelled by the momenta $p_x$ and $p_y$ of the colliding particles and satisfy

$$^0\langle p'_x, p'_y \text{ in}|p_x, p_y \text{ in}\rangle^0 = {}^0\langle p'_x|p_x\rangle^0 \, {}^0\langle p'_y|p_y\rangle^0 + {}^0\langle p'_x|p_y\rangle^0 \, {}^0\langle p'_y|p_x\rangle^0 \ .$$

The corresponding basis of true in-going scattering states is constructed by means of Bloch's linear operator $\mathrm{U}_2$ and found to be

$$\begin{aligned} |p_x, p_y \text{ in}\rangle &= |p_x, p_y \text{ in}\rangle^0 + \frac{\kappa \mathrm{i}}{2} \Bigg\{ \delta(p_x + p_y)|\Omega\rangle^0 \\ &\quad - \frac{1}{4} {\sum_{x,y,z}}' \mathrm{e}^{\mathrm{i}P(x+y)} \sin(|q||x-y|)|x,y,z,z+1\rangle^0 \Bigg\} + \mathcal{O}(\kappa^2) \ . \end{aligned}$$

In terms of these, normalizable in-going scattering states can be written as

$$|f_1, f_2 \text{ in}\rangle = \int_{-\pi}^{\pi} \mathrm{d}p_x \int_{-\pi}^{\pi} \mathrm{d}p_y \, \tilde{f}_1(p_x) \tilde{f}_2(p_y) |p_x, p_y \text{ in}\rangle \ . \tag{4.17}$$

where the one-particle wave functions $\tilde{f}_1$ and $\tilde{f}_2$ are supported in momentum space in such a way that the corresponding ranges of group velocities $\mathrm{d}E_1(p)/\mathrm{d}p$ have opposite sign. The physical characterization of $|f_1, f_2 \text{ in}\rangle$ is that for large negative times it evloves into a state which describes two widely seperated one-particle wave packets approaching each other. By an argumentation similar to the one applied in the case of localized one-particle states we expect that for any local observable $\mathbb{O}(x)$ the in-going scattering states $|f_1, f_2 \text{ in}\rangle$ should satisfy (cf. ref. [16])

$$\begin{aligned} \langle f_1, f_2 \text{ in}|\mathbb{O}(x)|f_1, f_2 \text{ in}\rangle &= \langle f_1|f_1\rangle\langle f_2|\mathbb{O}(x)|f_2\rangle + \langle f_2|f_2\rangle\langle f_1|\mathbb{O}(x)|f_1\rangle \\ &\quad - \langle f_1|f_1\rangle\langle f_2|f_2\rangle\langle\Omega|\mathbb{O}(x)|\Omega\rangle \ , \end{aligned} \tag{4.18}$$

up to terms which vanish rapidly when the distance between the wave packets $f_1$ and $f_2$ is made large. In appendix C it is shown that the states (4.17) indeed fullfil this basic physical property.

## 5 The $(3+1)$-dimensional model

The difficulty with the higher dimensional models is that we do not have an analytical solution for the Green's function (3.20). Therefore we will not be able to present the next-to-leading order approximation of the in-going scattering states in a closed form as for the simple $(1+1)$-dimensional case. However, due to the locality of the potential, to calculate the leading and next-to-leading order transition matrix the Green's function is needed only at a few points near the origin.

## 5.1 Small momentum expansion

Bloch's perturbation theory yields the two-particle energy $E_2(\boldsymbol{q})$ as a power series in $\kappa$ which is convergent if the inverse coupling is sufficiently small. We introduce momentum spherical variables

$$q_1 = q\sin(\theta)\cos(\phi)\ ,\quad q_2 = q\sin(\theta)\sin(\phi)\ ,\quad q_3 = q\cos(\theta)\ . \tag{5.1}$$

Since the momentum cartesian coordinates are restricted to the interval $q_k \in (-\pi, \pi]$, for $q \geq \pi$ the angular variables do not cover the whole range $0 \leq \theta \leq \pi$ and $0 \leq \phi \leq 2\pi$, respectively. Instead they must be restricted to some subdomain of the surface of the sphere since, pictorially speaking, the sphere of radius $q \geq \pi$ intersects the cube of edgelength $2\pi$. Physically the range $q \geq \pi$ is the region where the model is dominated by lattice artefacts. But if the abolute value of the relative momentum is less than $\pi$ the two-particle energy is a smooth function of $q$. It follows that $E_2(\boldsymbol{q})$ has a convergent series representation in powers of $q$. In the limit of small momentum the continuum rotational symmetry is restored on the lattice. Therefore the first non-constant contributions to this series is proportional to $q^2$ (cf. eq. (2.36)), hence

$$E_2(\boldsymbol{q}) = 2m + \frac{q^2}{m_*} + \sum_{\nu=3}^{\infty} b_\nu(\Omega;\kappa)\, q^\nu\ . \tag{5.2}$$

It is convenient to intoduce a modified absolute value of momentum

$$q_* \stackrel{\text{def}}{=} \sqrt{m_*\,\{E_2(q) - 2m\}}\ . \tag{5.3}$$

At low energy or small momentum, respectively, one has $q_* = q$, so (5.3) is a suitable parameter for on-shell small momentum expansions of the scattering quantities. From the implicit function theorem we conclude that eq. 5.2) can be inverted for small values of $q_*$. This leads to a series expansion for the (true) absolute value of the relative momentum of the scattering particles

$$q = q_* + \sum_{\nu=2}^{\infty} c_\nu(\Omega;\kappa)\, q_*^\nu\ . \tag{5.4}$$

The coefficients $c_\nu$ are polynomials of $\sin(\theta)$, $\sin(\phi)$ and $\cos(\theta)$, $\cos(\phi)$. Therefore they can be expressed as sums of spherical harmonics $Y_l^m(\Omega)$.

In order to diagonalize the scattering matrix one proceeds as in the $(1+1)$-dimensional case. We define free states $|E_2,\Omega\rangle$ and scattering states $|E_2,\Omega\ \text{in}\rangle$ through

$$|E_2,\Omega\rangle = q\left(\frac{\mathrm{d}E_2}{\mathrm{d}q}\right)^{-1/2}\Phi_\mathrm{s}(\boldsymbol{q})\ ,\quad |E_2,\Omega\ \text{in}\rangle = q\left(\frac{\mathrm{d}E_2}{\mathrm{d}q}\right)^{-1/2}\Phi_\mathrm{s}^{\text{in}}(\boldsymbol{q})\ . \tag{5.5}$$

From the normalization (3.16) it follows that

$$\langle E_2',\Omega' \mid E_2,\theta,\phi\rangle = \langle E_2',\Omega'\ \text{in} \mid E_2,\theta,\phi\ \text{in}\rangle = \left(\frac{\mathrm{d}E_2'}{\mathrm{d}q}\frac{\mathrm{d}E_2}{\mathrm{d}q}\right)^{-1/2}\delta(q'-q)\,\delta_\mathrm{s}(\Omega'-\Omega)\ , \tag{5.6}$$

where

$$\delta_\mathrm{s}(\Omega'-\Omega) = \frac{1}{2\sin(\theta)}\left\{\delta(\theta'-\theta)\delta(\phi'-\phi) + \delta(\theta'+\theta-\pi)\delta(\phi'-\phi-\pi)\right\}\ . \tag{5.7}$$

The r.h.s. of eq. (5.7) is zero unless the primed angular variables either coincide with the unprimed ones or are related to them by a space reflection. In both cases $E_2' = E_2$ if and only if $q' = q$. So can substitute

$$\delta(q'-q) = \frac{\mathrm{d}E_2}{\mathrm{d}q}\delta(E_2'-E_2) \tag{5.8}$$

on the r.h.s. of eq. (5.6) and end up with

$$\langle E_2', \Omega' \mid E_2, \Omega \rangle = \delta(E_2' - E_2)\,\delta_{\mathrm{s}}(\Omega' - \Omega) \ . \tag{5.9}$$

With respect to the new basis of states the scattering matrix is given by

$$\mathsf{S}(E_2', \Omega'; E_2, \Omega) = \mathsf{S}(E_2; \Omega', \Omega)\delta(E_2' - E_2) \ , \tag{5.10}$$

where

$$\mathsf{S}(E_2; \Omega', \Omega) = \delta_{\mathrm{s}}(\Omega' - \Omega) - 2\pi \mathrm{i} \kappa\, q q' \left( \frac{\mathrm{d}E_2}{\mathrm{d}q} \frac{\mathrm{d}E_2}{\mathrm{d}q'} \right)^{-1/2} \mathbb{T}(E_2; \Omega', \Omega) \ . \tag{5.11}$$

Remembering the results of section 3 we known that the transition matrix is determined by the Green's function at definite coordinate points plus some kinematical contributions. Furthermore, the integral representation (3.20) together with (5.2) allows us to expand $G(\boldsymbol{q}; \boldsymbol{r})$ as a convergent power series in $q$ if the relative momentum is sufficiently small. By means of (5.4) one then deduces a series representation for the scattering matrix at fixed energy

$$\mathsf{S}(E_2; \Omega', \Omega) = \delta_{\mathrm{s}}(\Omega' - \Omega) + \sum_{\nu=1}^{\infty} S_\nu(\Omega', \Omega)\, q_*^\nu \ . \tag{5.12}$$

The coefficients $S_\nu$ are sums of spherical harmonics in the momentum angular variables $(\Omega', \Omega)$, where the maximum angular momentum quantum number $l$ increases with increasing order in $q_*$. Eq. (5.12) is a general result independent of the order of truncation for the effective Hamiltonians. The calculation of the first few orders provides an insight into the small momentum behaviour of the model's scattering quantities. For instance, one will be able to determine the scattering length and get an idea of the angular dependence of the scattering matrix.

To be explicit we perform the small momentum expansion up to order $\mathcal{O}(q_*^5)$ using the results from the next-to-leading order approximations derived in the preceeding sections. The two-particle energy reads

$$E_2(\boldsymbol{q}) = 2m + \frac{\widehat{\boldsymbol{q}}^2}{m_*} - \frac{\kappa^2}{2} (\widehat{\boldsymbol{q}}^2)^2 \ , \tag{5.13}$$

where

$$m = 2 - 6\kappa(1 + \kappa) \ , \quad m_*^{-1} = 2\kappa(1 + 3\kappa) \ . \tag{5.14}$$

It is a specialty of this approximation that $E_2$ is a polynomial in $\widehat{q} = \sqrt{\widehat{\boldsymbol{q}}^2}$. In appendix C we argue that the Green's function has a convergent series representation

$$G(\boldsymbol{q}; \boldsymbol{r}) = \sum_{\nu=0}^{\infty} G_\nu(\boldsymbol{r})\, \widehat{q}^\nu \tag{5.15}$$

and calculate the coefficients up to the fifth order to 10 significant digits. The results for $\boldsymbol{r} = 0$, $\boldsymbol{r} = \hat{1}$ and some values of $\kappa$ are listed in tables 2 and 3 of appendix C. Using the definition (5.3) we obtain from (5.13)

$$\widehat{q} = q_* + \frac{1}{4} \kappa^2 m_*\, q_*^3 + \frac{7}{32} \kappa^4 m_*^2\, q_*^5 + \mathcal{O}(q_*^7) \ . \tag{5.16}$$

Together with (5.16) eq. (5.15) provides a small momentum expansion for the Green's function. To handle the remaining terms in eq. (5.11) we need to expand the absolute value of the relative momentum (cf. eq. (5.4))

$$q = q_* - \frac{1}{2} b_4 m_*\, q_*^3 + \left( \frac{7}{8} b_4^2 m_*^2 - \frac{1}{2} b_6 m_* \right) q_*^5 + \mathcal{O}(q_*^7) \ , \tag{5.17}$$

where

$$b_4 = -\frac{\kappa^2}{2} - \frac{\kappa}{6}(1 + 3\kappa) f_1(\Omega) \ , \quad b_6 = \frac{\kappa^2}{12} f_1(\Omega) + \frac{\kappa}{180}(1 + 3\kappa) f_2(\Omega) \ ,$$

and

$$f_1(\Omega) = \sin^4\theta\cos^4\phi + \sin^4\theta\sin^4\phi + \cos^4\theta \ , \tag{5.18}$$
$$f_2(\Omega) = \sin^6\theta\cos^6\phi + \sin^6\theta\sin^6\phi + \cos^6\theta \ . \tag{5.19}$$

Combining the above results the desired low energy expansion for the scattering matrix is now obtained by some lengthy but straightforward algebra. We state the final result

$$\mathsf{S}(E_2;\Omega',\Omega) = \delta_s(\Omega'-\Omega) + \sum_{\nu=1}^{5} S_\nu\, q_*^\nu + \mathcal{O}(q_*^6) \ , \tag{5.20}$$

where

$$S_1 = \frac{\alpha_{11}}{\pi^2} \ , \quad S_3 = \frac{\alpha_{31}+\alpha_{32}[f_1(\Omega')+f_1(\Omega)]}{\pi^2} \ , \quad S_2 = \frac{\alpha_{21}}{\pi^2} \ , \quad S_4 = \frac{\alpha_{41}+\alpha_{42}[f_1(\Omega')+f_1(\Omega)]}{\pi^2} \ , \tag{5.21}$$

and

$$S_5 = \frac{1}{\pi^2}\{\alpha_{51}+\alpha_{52}[f_1(\Omega')+f_1(\Omega)]+\alpha_{53}[f_1^2(\Omega')+f_1^2(\Omega)]+\alpha_{54}f_1(\Omega')f_1(\Omega) + \alpha_{55}[f_2(\Omega')+f_2(\Omega)]+\alpha_{56}f_3(\Omega',\Omega)\} \ . \tag{5.22}$$

The function $f_3$ appears when expanding the angular dependent part of the next-to-leading order transition matrix (3.50)

$$f_3(\Omega',\Omega) = \sin^2\theta'\sin^2\theta\cos^2\phi\cos^2\phi' + \sin^2\theta\sin^2\theta'\sin^2\phi\sin^2\phi' + \cos^2\theta\cos^2\theta' \ . \tag{5.23}$$

The coefficients $\alpha_{ij}$ are functions of $\kappa$ and $G_\nu(\boldsymbol{r})$, for instance

$$\alpha_{11} = \mathrm{i}\,\frac{m_*}{8}\,\frac{G_0(0) - 30\kappa\,\{G_0(0)-G_1(0)\}^2}{G_0(0)^2} \ . \tag{5.24}$$

The higher order contributions become more and more lengthy and are unsuitable for publication. If one substitues a certain value for the inverse coupling together with the results of tables 2 and 3 of appendix C they are just complex numbers.

For the diagonalization of the scattering matrix we expand (5.20) into spherical harmonics. Since we are confined to the subspace of symmetric wave functions we have to restrict ourselves to even $l$ values. Hence

$$\mathsf{S}(E_2;\Omega',\Omega) = \sum_{l'=0}^{\infty}\sum_{m'=-2l'}^{2l'}\sum_{l=0}^{\infty}\sum_{m=2l}^{2l} Y_{2l'}^{m'}(\Omega')^*\,\mathcal{S}(E_2)_{2l'\,m',2l\,m}\,Y_{2l}^{m}(\Omega) \ , \tag{5.25}$$

with

$$\mathcal{S}(E_2)_{2l'\,m',2l\,m} = \delta_{ll'}\delta_{mm'} + \sum_{\nu=1}^{5}\mathcal{S}^\nu_{2l'\,m',2l\,m}\,q_*^\nu + \mathcal{O}(q_*^6) \ , \tag{5.26}$$

and

$$\mathcal{S}^\nu_{2l'\,m',2l\,m} = \int \mathrm{d}\Omega' \int \mathrm{d}\Omega\, Y_{2l'}^{m'}(\Omega')\,S_\nu\,Y_{2l}^{m}(\Omega)^* \ . \tag{5.27}$$

Up to the second order (5.20) does not depend on the the angular variables. So the correction to the identity matrix in (5.26) is only for the first eigenvalue. As a consequence the scattering length can be determined straightforwardly

$$a_0 = \lim_{q_*\to 0}\frac{\mathcal{S}^1_{00,00}}{2\mathrm{i}} = \frac{2\alpha_{11}}{\mathrm{i}\pi} \ . \tag{5.28}$$

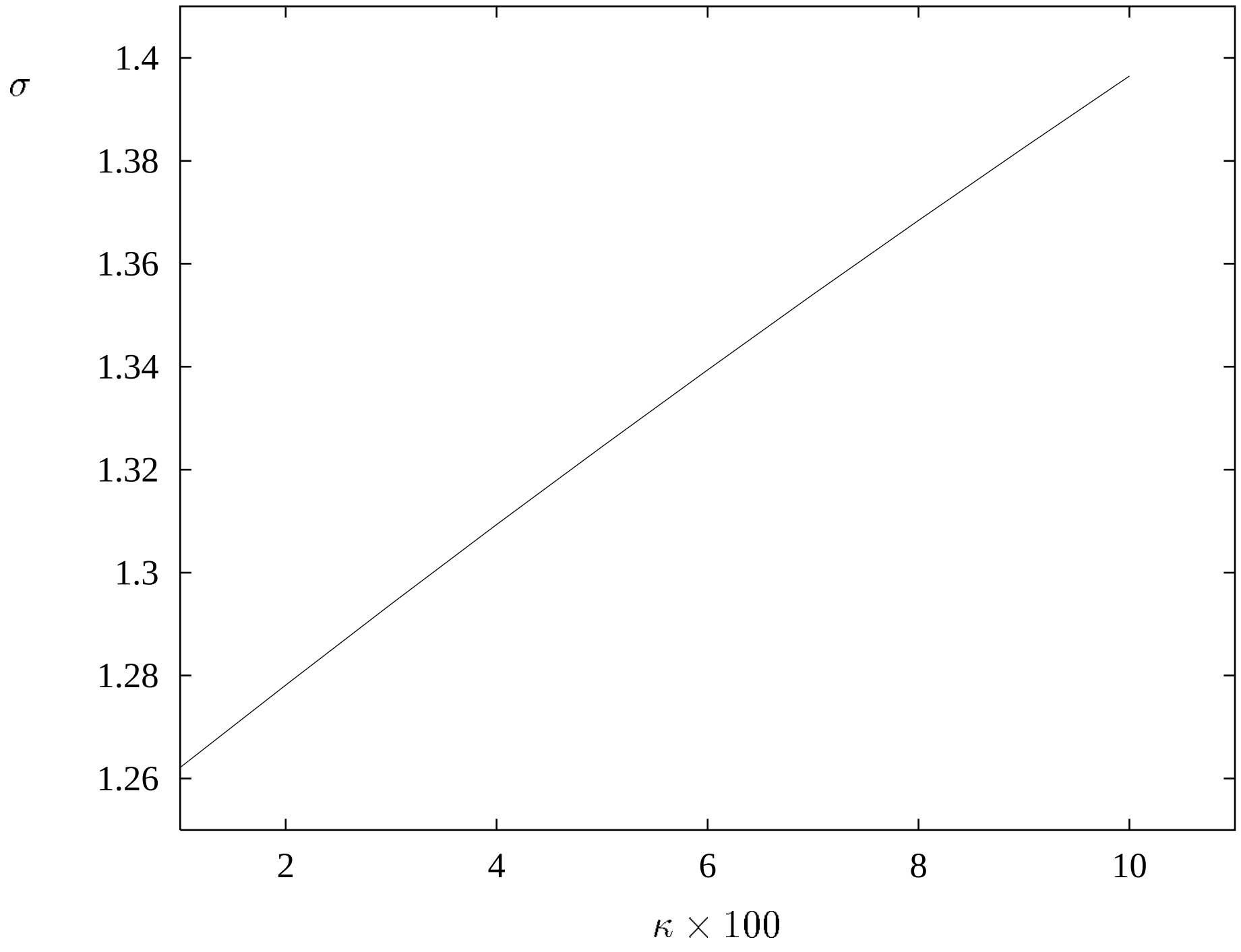

Figure 5: The geometric cross section as a function of the inverse coupling.

Fig. 1 shows the geometric cross section $\sigma = 4\pi a_0^2$ as a function of the inverse coupling $\kappa$. From (2.35) one obtains $\kappa_c = 0.264 + \mathcal{O}(\kappa^3)$ for the $(3+1)$-dimensional model. Therefore it is not reasonable to go beyond $\kappa = 0.1$ in a strong coupling expansion. The interaction strength increases almost linearly with increasing magnitude of $\kappa$. Nevertheless the sensitivity is not very strong since an increase of a factor of 10 only results in a change of 11 per cent for the cross section. The corresponding area is about 1.3 plaquettes on the lattice.

To evaluate the higher order contributions (5.27) we write the functions $f_i$ as sums of spherical harmonics and use the orthogonality relations. To third and fourth order the maximum value is $l = 4$ leading to a non-trivial $(15 \times 15)$ submatrix. If we include the fifth order contribution the maximum value increases to $l = 8$ and we obtain a $(45 \times 45)$ submatrix.

### 5.2 The phase shifts

The scattering matrix is invariant with respect to cubic transformations. As a consequence its eigenstates come as multiplets which transform according to the irreducible representations of the cubic group $\mathrm{O}(3, \mathbb{Z})$. It is well known that the special cubic group $\mathrm{SO}(3, \mathbb{Z})$ has 5 irreducible representations. They are denoted by $A_1$,$A_2$,$E$,$T_1$ and $T_2$, and their dimensionalities are 1,1,2,3 and 3, respectively. The irreducible representations of the full cubic group are characterized by one of the irreducible representations of the special cubic group and the parity $\pm 1$ which indicates the transformation behaviour of the wave functions under the space reflections $\boldsymbol{r} \to -\boldsymbol{r}$. These representations will be denoted by superscripts $A_1^+$, $A_1^-$ etc. Since the two-particle sector of the Ising model consists of symmetric wave functions we will be concerned with even parity representations only. Representations of the cubic group naturally arise from representations of the orthogonal group $\mathrm{O}(3)$. This can be illustrated by considering the vector space $\chi_l$ of all homogeneous polynomials of degree $l$ in $\boldsymbol{r} \in \mathbb{R}^3$ as the representation space of the $l$-dimensional irreducible representation of $\mathrm{O}(3)$. Through the obvious restriction, $\chi_l$ is also a representation space of the cubic group $\mathrm{O}(3, \mathbb{Z})$. The decomposition into irreducible representations can be easily found by analyzing the properties of the basic polynomials of $\chi_l$ under cubic

Table 1: The coefficients $\delta_\Gamma^{(\nu)}$.

| | $\kappa = 0.01$ | $\kappa = 0.05$ | $\kappa = 0.1$ |
|---|---|---|---|
| $\delta^{(1)}_{A_1^+}$ | $-0.3169$ | $-0.3247$ | $-0.3334$ |
| $\delta^{(2)}_{A_1^+}$ | $2.6\ 10^{-5}$ | $5.6\ 10^{-4}$ | $0.0019$ |
| $\delta^{(3)}_{A_1^+}$ | $-0.0141$ | $-0.0157$ | $-0.0184$ |
| $\delta^{(4)}_{A_1^+}$ | $1.7\ 10^{-5}$ | $4.1\ 10^{-4}$ | $0.0015$ |
| $\delta^{(5)}_{A_1^+}$ | $-0.0074$ | $-0.0033$ | $-7.2\ 10^{-4}$ |
| $\delta^{(5)}_{E^+}$ | $5.1\ 10^{-5}$ | $2.3\ 10^{-4}$ | $4.1\ 10^{-4}$ |

transformations. For $l = 0, 2, 4$ one gets

$$\begin{aligned} \mathbf{0} &= A_1^+ \ , \\ \mathbf{2} &= E^+ \oplus T_2^+ \\ \mathbf{4} &= A_1^+ \oplus E^+ \oplus T_1^+ \oplus T_2^+ \ . \end{aligned}$$

The eigenvalues are parametrized through phase shifts $\delta_\Gamma$ as $\exp(2i\delta_\Gamma)$, where the index $\Gamma$ denotes the corresponding irreducible representation of the cubic group. At small momentum we make the ansatz

$$\delta_\Gamma = \sum_{\nu=1}^{\infty} \delta_\Gamma^{(\nu)} q_*^\nu \ . \tag{5.29}$$

To calculate the coefficients $\delta_\Gamma^{(\nu)}$ up to the fifth order we use Bloch's perturbation theory. Since the scattering matrix is unitary but not hermitian we introduce

$$\mathcal{S}_r = \frac{1-\mathrm{i}}{2}\mathcal{S} + \frac{1-\mathrm{i}}{2}\mathcal{S}^\dagger = V + W q_*^3 \ , \tag{5.30}$$

where

$$V = \mathbb{1} + \mathcal{S}_r^1\, q_* + \mathcal{S}_r^2\, q_*^2 \ . \tag{5.31}$$

From (5.21) and (5.27) it follows that $V$ is diagonal. The non-trivial eigenvalue $v_1 = 1 + (4/\pi)\alpha_{11} q_* + (4/\pi)\alpha_{21} q_*^2$ has multiplicity one, so $v_1$ corresponds to the representation $A_1^+$. Using Bloch's formula (2.15) the higher order contributions can be determined straightforwardly

$$\cos(2\delta_{A_1^+}) + \sin(2\delta_{A_1^+}) = v_1 + W_{1,1}\, q_*^3 + \sum_{n=2}^{45} \frac{W_{1,n} W_{n,1}}{v_1 - 1}\, q_*^6 \tag{5.32}$$

To find out if there are more scattering channels at low energies we have to apply perturbation theory to the eigenspace corresponding to the infinitely degenerate eigenvalue $v_2 = 1$. Up to order $\mathcal{O}(q_*^5)$ this requires the diagonalization of the reduced matrix

$$(\mathcal{S}_r')_{i,j} = W_{i+1,j+1}\, q_*^3 + \frac{W_{i+1,1} W_{1,j+1}}{1 - v_1}\, q_*^6 \ . \tag{5.33}$$

It is easy to convince oneself that this $(44 \times 44)$ matrix is in fact of the order $\mathcal{O}(q_*^5)$. So it can be diagonalized numerically. As a result we find one non-trivial eigenvalue of multiplicity two which is therefore related to the two-dimensional irreducible representation $E^+$.

The resulting coefficients $\delta_\Gamma^{(\nu)}$ for several values of the inverse coupling $\kappa$ are collected in table 1. Since the effective potential is short range one expects the s-wave scattering to be dominant at low energy.

### 5.3 Beyond the small momentum regime

In general, the knowledge about the small momentum behaviour of the scattering quantities is not sufficient to answer more subtle questions such as the existence of resonances, for example. Therefore the investigation must be extended beyond the small momentum regime. This means that we have to find a diagonalization of the scattering matrix independent of the value of the two-particle energy. What makes life difficult is the fact that on the lattice there is no rotational symmetry. As a consequence there is no obvious choice of eigenstates for the scattering operator. Contrary to the small momentum expansion the methods will be different for different models and orders of truncation, and no general strategy can be presented.

For the present case, a promising procedure is suggested by the dispersion relation (5.13). Since the energy is a simple polynomial of $\widehat{q}$ it seems reasonable to introduce new momentum variables

$$\widehat{q}_k = 2\sin\left(\frac{q_k}{2}\right) \ , \quad -2 < \widehat{q}_k \leq 2 \ . \tag{5.34}$$

The plane waves and the (improper) scattering states which are related to $\Phi_{\rm s}(\boldsymbol{q})$ and $\Phi_{\rm s}^{\rm in}(\boldsymbol{q})$ via the coordinate transformation (5.34) are denoted by $\Phi_{\rm s}(\widehat{\boldsymbol{q}})$ and $\Phi_{\rm s}^{\rm in}(\widehat{\boldsymbol{q}})$, respectively. The normalization condition (3.16) translates to

$$(\Phi_{\rm s}(\widehat{\boldsymbol{q}}'), \Phi_{\rm s}(\widehat{\boldsymbol{q}})) = (\Phi_{\rm s}^{\rm in}(\widehat{\boldsymbol{q}}'), \Phi_{\rm s}^{\rm in}(\widehat{\boldsymbol{q}})) = \frac{Z(\widehat{\boldsymbol{q}})}{2}\{\delta(\widehat{\boldsymbol{q}}'-\widehat{\boldsymbol{q}}) + \delta(\widehat{\boldsymbol{q}}'+\widehat{\boldsymbol{q}})\} \ , \tag{5.35}$$

where the normalization function is given by

$$Z(\widehat{\boldsymbol{q}}) = \frac{1}{8}\sqrt{\left(4-\widehat{q}_1^2\right)\left(4-\widehat{q}_2^2\right)\left(4-\widehat{q}_3^2\right)} \ . \tag{5.36}$$

We make a further change of coordinates and introduce momentum spherical variables

$$\widehat{q}_1 = \widehat{q}\sin(\widehat{\theta})\cos(\widehat{\phi}) \ , \quad \widehat{q}_2 = \widehat{q}\sin(\widehat{\theta})\sin(\widehat{\phi}) \ , \quad \widehat{q}_3 = \widehat{q}\cos(\widehat{\theta}) \ . \tag{5.37}$$

Then eq. (5.35) reads

$$(\Phi_{\rm s}(\widehat{\boldsymbol{q}}'), \Phi_{\rm s}(\widehat{\boldsymbol{q}})) = \frac{4\kappa - 2\kappa^2(\widehat{q}^2-6)}{\widehat{q}} Z(E_2,\widehat{\Omega})\,\delta(E_2'-E_2)\,\delta_{\rm s}(\Omega'-\Omega) \ , \tag{5.38}$$

where

$$\widehat{q} = \left\{\frac{1-\sqrt{1-2\kappa^2 m_*^2(E_2-2m)}}{\kappa^2 m_*}\right\}^{1/2} \tag{5.39}$$

and the normalization function becomes

$$Z(E_2,\widehat{\Omega}) = \begin{cases} Z\left(\widehat{\boldsymbol{q}}\left(\widehat{q},\widehat{\theta},\widehat{\phi}\right)\right) & \text{if } -2 < \widehat{q}_k\left(\widehat{q},\widehat{\theta},\widehat{\phi}\right) \leq 2 \ , \\ 0 & \text{otherwise} \ . \end{cases} \tag{5.40}$$

The next step is to expand the free states $\Phi_{\rm s}(\widehat{\boldsymbol{q}})$ and the scattering states $\Phi_{\rm s}^{\rm in}(\widehat{\boldsymbol{q}})$ into spherical harmonics

$$\Phi_{\rm s}(\widehat{\boldsymbol{q}}) = \sum_{l=0}^{\infty}\sum_{m=-2l}^{2l} Y_{2l}^m(\widehat{\Omega})\,|2l\,m;E_2\rangle \ , \tag{5.41}$$

$$\Phi_{\rm s}^{\rm in}(\widehat{\boldsymbol{q}}) = \sum_{l=0}^{\infty}\sum_{m=-2l}^{2l} Y_{2l}^m(\widehat{\Omega})\,|2l\,m;E_2\ {\rm in}\rangle \ . \tag{5.42}$$

The new basis vectors are *not* orthonormal. Instead, from (5.38) it follows that

$$\langle 2l'\,m';E_2' \mid 2l\,m;E_2\rangle = \langle 2l'\,m';E_2'\ {\rm in} \mid 2l\,m;E_2\ {\rm in}\rangle = \kappa\,\mathcal{Z}(E_2)_{2l'm',2lm}\,\delta(E_2'-E_2) \tag{5.43}$$

where we have introduced the normalization matrix

$$\mathcal{Z}(E_2)_{2l'\,m',2l\,m} = \frac{4-2\kappa(\widehat{q}^2-6)}{\widehat{q}} \int \mathrm{d}\widehat{\Omega}\, Z(E_2,\widehat{\Omega})\, Y_{2l'}^{m'}(\widehat{\Omega})\, Y_{2l}^{m}(\widehat{\Omega})^* \ . \tag{5.44}$$

Starting from eq. (3.11) and using the above formulae it is an easy task to expand the scattering matrix

$$\mathsf{S}(\widehat{\boldsymbol{q}}',\widehat{\boldsymbol{q}}) = \delta(E_2'-E_2)\sum_{l'=0}^{\infty}\sum_{m'=-2l'}^{2l'}\sum_{l=0}^{\infty}\sum_{m=2l}^{2l} Y_{2l'}^{m'}(\widehat{\Omega}')^*\, \mathcal{S}(E_2)_{2l'\,m',2l\,m}\, Y_{2l}^{m}(\widehat{\Omega}) \ , \tag{5.45}$$

with

$$\mathcal{S}(E_2)_{2l'\,m',2l\,m} = \kappa\, \mathcal{Z}(E_2)_{2l'\,m',2l\,m} - 2\pi\mathrm{i}\kappa\, \mathcal{T}(E_2)_{2l'\,m',2l\,m} \tag{5.46}$$

and

$$\mathcal{T}(E_2)_{2l'\,m',2l\,m} = \int \mathrm{d}\widehat{\Omega}' \int \mathrm{d}\widehat{\Omega}\, Y_{2l'}^{m'}(\widehat{\Omega}')\, \mathbb{T}(\widehat{\boldsymbol{q}}',\widehat{\boldsymbol{q}})\, Y_{2l}^{m}(\widehat{\Omega})^* \ . \tag{5.47}$$

In order to calculate the phase shifts it remains to diagonalize (5.46), i.e. we have to solve the generalized eigenvalue problem

$$\mathcal{S}v = s\mathcal{Z}v \ . \tag{5.48}$$

Since the normalization matrix (5.44) is invertible, the solution of the eigenvalue equation (5.48) for the matrices $\mathcal{S}$ and $\mathcal{Z}$ is equivalent to the diagonalization of the product

$$\mathcal{S}'(E_2) \stackrel{\mathrm{def}}{=} [\kappa\mathcal{Z}(E_2)]^{-1}\, \mathcal{S}(E_2) = \mathbb{1} - 2\pi\mathrm{i}\, \mathcal{Z}^{-1}(E_2)\, \mathcal{T}(E_2) \ . \tag{5.49}$$

Thanks to the well-suited momentum variables the determination of the transition matrix (5.47) is comparatively simple. Since the momentum dependence of the Green's function is via $\widehat{q}$ only, it is just the first part of (3.50) that varies with the angular variables $(\Omega',\Omega)$. We execute the transformation to the momentum spherical variables and express the result in terms of spherical harmonics

$$\begin{aligned}\sum_{k=1}^{2}\cos(q_k)\cos(q_k') &= \frac{1}{12}(\widehat{q}^2-6)^2 + \widehat{q}^4\,\frac{2\pi}{15}\, Y_2^0(\widehat{\Omega})Y_2^0(\widehat{\Omega}') \\ &\quad + \widehat{q}^4\,\frac{\pi}{15}\left\{Y_2^2(\widehat{\Omega}) + Y_2^{-2}(\widehat{\Omega})\right\}\left\{Y_2^2(\widehat{\Omega}') + Y_2^{-2}(\widehat{\Omega}')\right\} \ .\end{aligned} \tag{5.50}$$

Using the orthogonality relations the on-shell transition matrix can now be determined straightforwardly. One obtains

$$\begin{aligned}\mathcal{T}(E_2)_{2l'\,m',2l\,m} &= \frac{\beta(\widehat{q})}{2\pi^2}\,\delta_{l'0}\,\delta_{m'0}\,\delta_{l0}\,\delta_{m0} \\ &\quad -\widehat{q}^4\,\frac{\kappa}{60\pi^2}\{2\delta_{l'1}\,\delta_{m'0}\,\delta_{l1}\,\delta_{m0} + \delta_{l'1}\,\delta_{l1}\,(\delta_{m'2}-\delta_{m'-2})(\delta_{m2}-\delta_{m-2})\} \ ,\end{aligned} \tag{5.51}$$

where

$$\beta(\widehat{q}) = -G(\widehat{q};0)^{-1} + \frac{5}{6}\kappa\left[\widehat{q}^2 - 6 + 6\frac{G(\widehat{q};1)}{G(\widehat{q};0)}\right]^2 \ .$$

Finally we substite (5.51) into eq. (5.49) and obtain the scattering matrix

$$\mathcal{S}'(E_2) = \mathbb{1} - \frac{\mathrm{i}}{\pi}\left\{\beta(\widehat{q})\mathcal{S}_1'(E_2) - \widehat{q}^4\,\frac{\kappa}{30}\mathcal{S}_2'(E_2)\right\} \ , \tag{5.52}$$

where

$$\begin{aligned}\mathcal{S}_1'(E_2)_{2l'\,l,2l\,m} &= \mathcal{Z}^{-1}(E_2)_{2l'\,m',0\,0}\,\delta_{l0}\,\delta_{m0} \\ \mathcal{S}_2'(E_2)_{2l'\,l,2l\,m} &= 2\mathcal{Z}^{-1}(E_2)_{2l'\,m',2\,0}\,\delta_{l1}\,\delta_{m0} \\ &\quad + \left[\mathcal{Z}^{-1}(E_2)_{2l'\,m',2\,2} + \mathcal{Z}^{-1}(E_2)_{2l'\,m',2\,-2}\right]\,\left[\delta_{l1}\delta_{m2} + \delta_{l1}\delta_{m-2}\right] \ .\end{aligned}$$

In order to calculate the inverse of the normalization matrix (5.44) one has to distinguish between two cases. If $\widehat{q} < 2$ the integration in (5.44) is over the whole surface of the sphere. Using the orthogonality of the spherical harmonics it is straightforward to show that

$$\mathcal{Z}^{-1}(E_2)_{2l'\,m',2l\,m} = \frac{\widehat{q}}{4-2\kappa(\widehat{q}^2-6)} \int \mathrm{d}\widehat{\Omega}\, Z^{-1}(E_2,\widehat{\Omega})\, Y_{2l'}^{m'}(\widehat{\Omega})\, Y_{2l}^{m}(\widehat{\Omega})^* \ . \tag{5.53}$$

On the other hand, if $\widehat{q} \geq 2$, the multiplication of (5.44) and (5.53) does not lead to the identity matrix since the integration is restricted corresponding to (5.40) and the orthogonality relation does not apply. In this case we have to calculate $\mathcal{Z}$ using eq. (5.44) and invert it afterwards. Because the matrices are infinite this procedure requires a truncation of the basis $\{|2l\,m; E_2\rangle\}$ keeping only a finite number of elements. For simplicity and since I am not so much interested in lattice artefacts I restrict attention to the region $\widehat{q} < 2$ in the following. The inverse normalization function can be expanded in powers of the momentum square $\widehat{q}^2$

$$Z^{-1}(E_2,\widehat{\Omega}) = \sum_{\nu=0}^{\infty} \zeta_\nu(\widehat{\Omega}) \left(\widehat{q}^2\right)^\nu \ . \tag{5.54}$$

The coefficients $\zeta_\nu(\widehat{\Omega})$ are trigonometric functions of the angular coordinates $(\theta,\phi)$ and can be expressed through spherical harmonics. So the r.h.s. of eq. (5.53) is a sum of integrals over products of three spherical harmonics. From the representation theory of the rotation group we know that such integrals are related to Wigner's $3j$-symbols through (cf. ref. [17], for example)

$$\int \mathrm{d}\widehat{\Omega}\, Y_{l_1}^{m_1}(\widehat{\Omega}) Y_{l_2}^{m_2}(\widehat{\Omega}) Y_{l_3}^{m_3}(\widehat{\Omega}) = \sqrt{\frac{(2l_1+1)(2l_2+1)(2l_3+1)}{4\pi}} \begin{pmatrix} l_1 & l_2 & l_3 \\ 0 & 0 & 0 \end{pmatrix} \begin{pmatrix} l_1 & l_2 & l_3 \\ m_1 & m_2 & m_3 \end{pmatrix} \ .$$

Using the algebraic manipulation programme MAPLE [23] the series expansion (5.54) and the subsequent decomposition of the coefficients $\zeta_n(\widehat{\Omega})$ into spherical harmonics can be automated completely. The non-vanishing $3j$-symbols are implemented through Wigner's formula [24]

$$\begin{aligned} \begin{pmatrix} l_1 & l_2 & l_3 \\ m_1 & m_2 & m_3 \end{pmatrix} &= (-1)^{l_1-l_2-m_3} \sqrt{\frac{(l_3+l_1-l_2)!(l_3-l_1+l_2)!(l_1+l_2-l_3)!(l_3-m_3)!(l_3+m_3)!}{(l_1+l_2+l_3+1)!(l_1-m_1)!(l_1+m_1)!(l_2-m_2)!(l_2+m_2)!}} \\ &\quad \times \sum_{k=k_1}^{k_2} \frac{(-1)^{k+l_2+m_2}(l_2+l_3+m_1-k)!(l_1-m_1+k)!}{k!(l_3-l_1+l_2-k)!(l_3-m_3-k)!(l_1-l_2+m_3+k)!} \ , \end{aligned}$$

where $k_1 = \max\{0, l_1-l_2-m_3\}$ and $k_2 = \min\{l_3-m_3, l_3+l_2-l_1\}$ . Obviously, the more the momentum absolute value approaches $\widehat{q} = 2$, where $Z^{-1}(E_2,\widehat{\Omega})$ is singular, the more terms are needed in the expansion (5.54). Here we decided to go as far as $\widehat{q} = 1.8$ which already requires an order of truncation of 20 to guarantee a relative accuracy of 1 percent. The CPU time needed was a few hours on an HP 9000/735 RISC workstation.

Since the correction to the identity in (5.52) is zero except for the first six columns the diagonalization of the infinite scattering matrix $\mathcal{S}'(E_2)$ is reduced to the diagonalization of the upper left $(6\times 6)$ submatrix of (5.52). As for the small momentum regime we only find two non-trivial eigenvalues of multiplicity one and two corresponding to the scattering phases $\delta_{A_1^+}$ and $\delta_{E^+}$. The results for $\kappa = 1/100$ and $\kappa = 1/10$ as a function of the momentum

$$q_* = \sqrt{m_*(E_2-2m)} = \widehat{q}\sqrt{1-\frac{1}{2}\kappa^2 m_* \widehat{q}^2} \ . \tag{5.55}$$

are plotted in fig. 6. Further scattering channels will only appear if the range of the potential increases, i.e. if one includes the higher order contributions to the effective two-particle Hamiltonian.

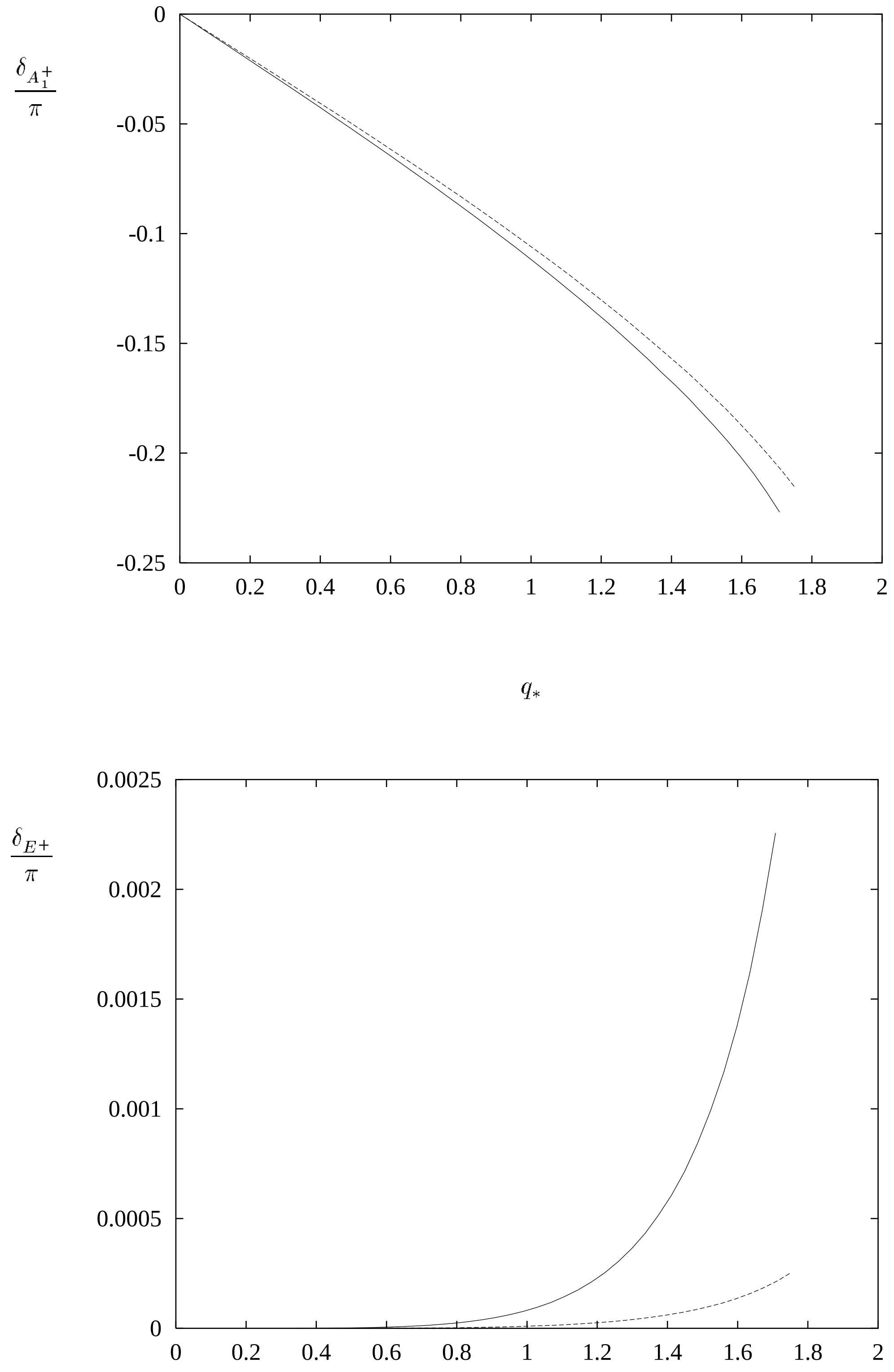

Figure 6: The scattering phases as a function of the momentum $q_*$ for $\kappa = 1/100$ (dashed lines) and $\kappa = 1/10$ (straight lines). The numerical error is less than 1 per cent for each value.

# 6 Concluding remarks

The purpose of this paper was to establish a technique to obtain strong coupling expansions for scattering phase shifts in Hamiltonian lattice field theories. The method was developed by means of the $(d+1)$-dimensional Ising model. Starting from Bloch's elegant perturbation theory for degenerate levels I derived a convergent series representation for the scattering states and the transition matrix. To demonstrate the feasibility of the approach the next-to-leading order approximation for the phase shifts in $d=1$ and $d=3$ spatial dimensions was computed.

The generalization of the present work to higher orders of the inverse coupling $\kappa$ is straightforward although the complexity of the effective two-particle Hamiltonian will rapidly increase. A further study of the Ising model might be interesting for its own sake, but I think that it is more promising to proceed in another direction. As alredady pointed out in the introduction the lattice Hamiltonian of pure gauge theory and QCD has the same general structure (1.1) as the model discussed here. Of course I am aware that these theories are much more involved than my test case. In particular the static two-particle spectrum is more extensive and will lead to more complicated effective Hamiltonians. But hopefully, these technical problems do not mean an insurmountable obstacle towards a qualitative insight into the low energy dynamics of hadrons.

*Acknowledgements:* The author whishes to thank Martin Lüscher for valuable and stimulating discussions on the subject, his constant interest in this work and a critical reading of the manuscript. Financial support by Studienstiftung des deutschen Volkes is gratefully acknowledged.

# Appendix A

In this appendix condition (2.41) is checked for the normalizable one-particle states (2.40). To show that they are well localized we verify (2.41) in the limit $|x| \to \infty$ to leading and next-to-leading order for the simplest local hermitean operators of the model. The generalization to more complicated, composite local observables is staightforward.

Substituting the explicit formulae for $|f\rangle$ and the ground state (2.21), a straightforward calculation leads to

$$\begin{aligned} \langle f|\mathbb{O}(x)|f\rangle &= (2\pi)^d \sum_{y,z} f^*(z) f(y) \Big\{ {}^0\langle z|\mathbb{O}(x)|y\rangle^0 \\ &\quad + \frac{\kappa}{4}\sum_{k=1}^{d}\sum_{z'}{}' \left( {}^0\langle z, z', z'+\hat{k}|\mathbb{O}(x)|y\rangle^0 + \text{c.c.}\right) \Big\} + \mathcal{O}(\kappa^2) \end{aligned} \tag{A.1}$$

for the l.h.s. and

$$\begin{aligned} \langle f|f\rangle\langle\Omega|\mathbb{O}(x)|\Omega\rangle &= (2\pi)^d \|f\|^2 \Big\{ {}^0\langle\Omega|\mathbb{O}(x)|\Omega\rangle^0 \\ &\quad + \frac{\kappa}{4}\sum_{k=1}^{d}\sum_{y} \left( {}^0\langle y, y+\hat{k}|\mathbb{O}(x)|\Omega\rangle^0 + \text{c.c.}\right) \Big\} + \mathcal{O}(\kappa^2)\ . \end{aligned} \tag{A.2}$$

for the r.h.s.

In case that the fundamental operator $\sigma_1(x)$ is chosen as local observable, the leading and next-to-leading order contributions to both (A.1) and (A.2) vanish. This is because $\sigma_1(x)|\Omega\rangle^0 \in \mathcal{E}_1$ and $\sigma_1(x)|y\rangle^0 \in \{\mathcal{E}_0 \cup \mathcal{E}_2\}$, so the matrix elements match eigenstates from different eigenspaces of $\mathbb{H}_0$.

Next we consider $\mathbb{O}(x) = \sigma_3(x)$. We remember $\sigma_3(x)|\Omega\rangle^0 = |\Omega\rangle^0$ as the defining relation for the static vacuum and recognize that $\sigma_2(x)|\Omega\rangle^0 = \mathrm{i}|x\rangle^0$. Using the commutation relations of the Pauli

matrices one obtains

$$\sigma_3(x)|y\rangle^0 = \sigma_3(x)\sigma_1(y)|\Omega\rangle^0 = \{1 - 2\delta(x-y)\}|y\rangle^0 \ . \tag{A.3}$$

Therefore the next-to-leading order contribution to (A.1) as well as (A.2) is zero. It remains to calculate the leading order of (A.1)

$$\langle f|\sigma_3(x)|f\rangle = (2\pi)^d \left(||f||^2 - 2|f(x)|^2\right) + \mathcal{O}(\kappa^2) \ . \tag{A.4}$$

Since $f(x)$ vanishes in the limit $|x| \to \infty$, the localization condition (2.41) is satisfied.

Finally we want to discuss a less simple situation and choose

$$\mathbb{O}(x) = \sum_{k=1}^{d} \sigma_1(x)\sigma_1(x+\hat{k}) \ .$$

Here the leading order contribution to (A.2) is zero because $\mathbb{O}(x)|\Omega\rangle^0 \in \mathcal{E}_2$, and the sum over the first order matrix elements yields $2d$. The evaluation of (A.1) is also straightforward and results in

$$\langle f|\mathbb{O}(x)|f\rangle = \frac{\kappa}{2} d\,(2\pi)^d\,||f||^2 + (2\pi)^d \sum_{k=1}^{d} \left\{ R_0(x;\hat{k}) + \frac{\kappa}{4} R_1(x;\hat{k}) \right\} + \mathcal{O}(\kappa^2) \ , \tag{A.5}$$

where the local expressions

$$\begin{aligned}
R_0(x;\hat{k}) &= f^*(x+\hat{k})f(x) + \text{ c.c.} \\
R_1(x;\hat{k}) &= \sum_{l=1}^{d} \left\{ f(x) \left[ f^*(x+\hat{k}+\hat{l}) + f^*(x+\hat{k}-\hat{l}) \right] + f(x+\hat{k}) \left[ f^*(x+\hat{l}) + f^*(x-\hat{l}) \right] \right\} \\
&\quad - f(x)f^*(x) - f(x+\hat{k})f^*(x+\hat{k}) + \text{ c.c.}
\end{aligned}$$

vanish in the limit $|x| \to \infty$.

# Appendix B

We show that the in-going scattering states (4.17) of the $(1+1)$-dimensional model satisfy condition (4.18) for the fundamental operators $\mathbb{O}(x) = \sigma_1(x)$ and $\mathbb{O}(x) = \sigma_3(x)$.

The r.h.s. of (4.18) is straightforwardly evaluated using the results of appendix A where the localization property of the model's one-particle states was established. So for $\mathbb{O}(x) = \sigma_1(x)$ the leading and next-to-leading order contributions vanish. From the explicit form of (4.17) it is easy to see that this is also true for the l.h.s. since $\sigma_1(x)|y,z\rangle^0 \in \{\mathcal{E}_1 \cup \mathcal{E}_3\}$, i.e. all matrix elements that occur match states from different satic eigenspaces.

In case of $\mathbb{O}(x) = \sigma_3(x)$ one finds

$$\text{r.h.s.} = 4\pi^2 ||f_1||^2 ||f_2||^2 - 8\pi^2 \left\{ |f_1(x)|^2 ||f_2||^2 + |f_2(x)|^2 ||f_1||^2 \right\} + \mathcal{O}(\kappa^2) \ . \tag{B.1}$$

For the calculation of the l.h.s. we remark that

$$\sigma_3(x)|y,z\rangle^0 = \{1 - 2\delta(x-y) - 2\delta(x-z)\}|y,z\rangle^0 \ .$$

Therefore the next-to-leading order contributions vanish and the result is

$$\begin{aligned}
\text{l.h.s.} \quad = \quad & \frac{1}{2\pi^2} \int_{-\pi}^{\pi} \mathrm{d}p'_z\, \tilde{f}_2(p'_z)^* \ldots \int_{-\pi}^{\pi} \mathrm{d}p_y\, \tilde{f}_1(p_y) \left\{ \sum_{y,z} \mathrm{e}^{\mathrm{i}(P-P')(y+z)} \sin(|q||y-z|) \sin(|q'||y-z|) \right. \\
& \left. -4 \sum_{y} \mathrm{e}^{\mathrm{i}(P-P')(x+y)} \sin(|q||x-y|) \sin(|q'||x-y|) \right\} + \mathcal{O}(\kappa^2) \ .
\end{aligned} \tag{B.2}$$

We fix the wave functions $\tilde{f}_1(p_x)$ and $\tilde{f}_2(p_y)$ such that they are supported around momentum values of different sign, say $p_x > 0$ and $p_y < 0$. Since

$$v \stackrel{\text{def}}{=} \frac{\mathrm{d}E_1(p)}{\mathrm{d}p} = 2\kappa\sin(p) + \kappa^2\sin(2p) + \mathcal{O}(\kappa^3)\ .$$

the corresponding group velocities are of different sign too, and their ranges do not overlap. It follows that $q > 0$ and $q' > 0$ in (B.2). So we can forget about the absolute values and write

$$\begin{aligned} \mathrm{e}^{\mathrm{i}(P-P')(y+z)}\sin[q(y-z)]\sin[q'(y-z)] = \tfrac{1}{4}\Big\{&\mathrm{e}^{\mathrm{i}[y(p_y-p'_y)+z(p_z-p'_z)]} \\ &+\mathrm{e}^{\mathrm{i}[y(p_z-p'_z)+z(p_y-p'_y)]} - \mathrm{e}^{\mathrm{i}[y(p_y-p'_z)+z(p_z-p'_y)]} - \mathrm{e}^{\mathrm{i}[y(p_z-p'_y)+z(p_y-p'_z)]}\Big\}\ . \end{aligned} \qquad \text{(B.3)}$$

Now the summation can be executed trivially. The momentum $\delta$-functions that result from the last two terms of (B.3) do not coincide with the supports of $\tilde{f}_1$ and $\tilde{f}_2$ though. Therefore they do not contribute to the subsequent $p$-integration. Carrying out these steps we find that the final result is equal to (B.1).

In a similar way an explicit check of condition (4.18) in case of $\mathbb{O}(x) = \sigma_1(x)\sigma_1(x+1)$ is possible, but requires considerably more elaborate calculations.

# Appendix C

We discuss the calculation of the three-dimensional Green's function. Our starting point is the integral representation

$$G(\boldsymbol{q};\boldsymbol{r}) = \int_{\mathcal{B}} \frac{\mathrm{d}^3p}{(2\pi)^3}\,\frac{\exp(\mathrm{i}\boldsymbol{pr})}{\epsilon_2(\boldsymbol{q}) - \epsilon_2(\boldsymbol{p}) + \mathrm{i}\rho}\ . \qquad \text{(C.1)}$$

The next-to-leading order approximation of the two-particle energy is a polynomial in $\hat{q}$

$$\epsilon_2(\boldsymbol{q}) = -12(1+\kappa) + 2(1+3\kappa)\hat{q}^2 - \frac{\kappa}{2}\hat{q}^4\ .$$

For $\kappa < \kappa_c$ the denominator of (C.1) factorizes as follows

$$\epsilon_2(\boldsymbol{q}) - \epsilon_2(\boldsymbol{p}) = -\frac{\kappa}{2}\left(\hat{q}^2 - \hat{p}^2 + \mathrm{i}\rho\right)\left(\hat{q}^2 + \hat{p}^2 - 12 - \frac{4}{\kappa}\right)\ .$$

As a consequence we can decompose the Green's function in two integrals of the same structure

$$G(\boldsymbol{q};\boldsymbol{r}) = \frac{G_1(\boldsymbol{q};\boldsymbol{r}) + G_2(\boldsymbol{q};\boldsymbol{r})}{2 - \kappa(\hat{q}^2 - 6)}\ , \qquad \text{(C.2)}$$

where

$$G_1(\boldsymbol{q};\boldsymbol{r}) = \int_{\mathcal{B}} \frac{\mathrm{d}^3p}{(2\pi)^3}\,\frac{\exp(\mathrm{i}\boldsymbol{pr})}{\hat{q}^2 - \hat{p}^2 + \mathrm{i}\rho}\ , \qquad \text{(C.3)}$$

$$G_2(\boldsymbol{q};\boldsymbol{r}) = \int_{\mathcal{B}} \frac{\mathrm{d}^3p}{(2\pi)^3}\,\frac{\exp(\mathrm{i}\boldsymbol{pr})}{\hat{q}^2 + \hat{p}^2 - 12 - 4/\kappa}\ . \qquad \text{(C.4)}$$

The first step towards the calculation of these three-dimensional integrals is to transform them into one-dimensional ones. To begin with we write

$$G_1(\boldsymbol{q};\boldsymbol{r}) = -\frac{\mathrm{i}}{2}\int_0^\infty \mathrm{d}t\,\exp\left\{\frac{\mathrm{i}}{2}t\left(\hat{q}^2 - 6\right)\right\}\mathrm{e}^{-t\rho}\prod_{k=1}^{3}\int_{-\pi}^{\pi}\frac{\mathrm{d}p_k}{2\pi}\exp(\mathrm{i}p_k r_k)\exp(\mathrm{i}t\cos p_k)\ . \qquad \text{(C.5)}$$

The reader may verify that (C.5) is indeed equal to (C.3) by simply executing the $t$-integration. Ref. [20, sect. 8.411] tells us that the integrals over the momentum components represent Bessel functions of the first kind

$$G_1(\boldsymbol{q};\boldsymbol{r}) = -\frac{\mathrm{i}}{2}\int_0^\infty \mathrm{d}t \exp\left\{\frac{\mathrm{i}}{2}t\left(\hat{q}^2-6\right)\right\}\prod_{k=1}^{3}\mathrm{i}^{r_k}J_{r_k}(t) \stackrel{\text{def}}{=} \int_0^\infty \mathrm{d}tF_1(t)\ . \tag{C.6}$$

The $t$-integration and the limit $\rho\to 0$ could be interchanged here because the integrand vanishes as $t^{-3/2}$ for $t\to\infty$ such that the integral (C.6) is well defined. Obviously such a heat kernel representation can also be deduced for (C.4)

$$G_2(\boldsymbol{q};\boldsymbol{r}) = \frac{\mathrm{i}}{2}\int_0^\infty \mathrm{d}t \exp\left\{-\frac{\mathrm{i}}{2}t\left(\hat{q}^2-6-\frac{4}{\kappa}\right)\right\}\prod_{k=1}^{d}\mathrm{i}^{r_k}J_{r_k}(t) \stackrel{\text{def}}{=} \int_0^\infty \mathrm{d}tF_2(t)\ . \tag{C.7}$$

The Bessel functions have asymptotic expansions for large arguments [20, sect. 8.451]. This leads to a corresponding expansion for the above functions $F_1(t)$ and $F_2(t)$ $F_j(t), j=1,2$

$$F_j(t) = (-1)^j\,\frac{\mathrm{i}}{2}\sum_{\mu=0}^{3}\sum_{\nu=0}^{\infty} b_{\mu,\nu}(\boldsymbol{r})\mathrm{e}^{-tz_j}t^{-\nu-3/2}\ ,\quad j=1,2\ , \tag{C.8}$$

where

$$z_1 = \frac{\mathrm{i}}{2}(12-4\mu-\hat{q}^2)\ ,\quad z_2 = \frac{\mathrm{i}}{2}(\hat{q}^2-4\mu-4/\kappa)\ . \tag{C.9}$$

The $b_{\mu,\nu}(\boldsymbol{r})$ are analytic functions of the components $r_k$ and the series is rapidly convergent for large values of $t$.

Next we decompose the integrals (C.6) and (C.7) in three parts

$$G_j(\boldsymbol{q};\boldsymbol{r}) = \int_0^{t_N}\mathrm{d}tF_j(t) + \int_{t_N}^\infty \mathrm{d}tF_j^N(t) + \int_{t_N}^\infty \mathrm{d}t\{F_j(t)-F_j^N(t)\}\ . \tag{C.10}$$

Here $F_j^N(t)$ denotes the asymptotic expansion (C.8) truncated after $\nu=N$ and $t_N$ is some positive number. If we substitute (C.8) the integrand of the third contribution to (C.10) is just the rest of the asymptotic expansion. One can estimate

$$|F_j(t)-F_j^N(t)| < \frac{1}{2}\,|t|^{-N-5/2}\sum_{\mu=0}^{3}|b_{\mu,N+1}(\boldsymbol{r})|\ ,$$

hence

$$\left|\int_{t_N}^\infty \mathrm{d}t\{F_j(t)-F_j^N(t)\}\right| < \frac{1}{2}\,t_N^{-N-3/2}\sum_{\mu=0}^{3}|b_{\mu,N+1}(\boldsymbol{r})|\ . \tag{C.11}$$

We demand that (C.11) is smaller than some error $\varepsilon$ and choose some order of truncation $N$. For any $\boldsymbol{r}$ eq. (C.11) then fixes the value of $t_N$ and one obtains an approximation of the Green's function where the error is perfectly under control

$$G_j(\boldsymbol{q};\boldsymbol{r}) = \int_0^{t_N}\mathrm{d}tF_j(t) + \int_{t_N}^\infty \mathrm{d}tF_j^N(t) + \mathcal{O}(\varepsilon)\ . \tag{C.12}$$

To guarantee an accurracy of 10 digits I chose $\varepsilon = 10^{-12}$. The order of truncation was taken to be $N=15$ which implies $t_N=12$ for $\boldsymbol{r}=0$ and $t_N=11$ for $\boldsymbol{r}=\hat{1}$.

In the following we will first describe how the representation (C.12) is used to calculate the Green's function for any given values $(\boldsymbol{q};\boldsymbol{r})$ and then discuss the series expansion of (C.12) in powers of $\hat{q}$.

Making use of [20, sect. 3.381] the infinite integral contributing to (C.12) can be expressed as a sum of incomplete Gamma functions

$$\int_{t_N}^{\infty} \mathrm{d}t F_j^N(t) = (-1)^j \frac{\mathrm{i}}{2} \sum_{\mu=0}^{3} \sum_{\nu=0}^{N} b_{\mu,\nu}(\boldsymbol{r}) \begin{cases} \left[(\nu+1/2)\, t_N^{\nu+1/2}\right]^{-1} & \text{if } z_j = 0 \\ z_j^{\nu+1/2}\, \Gamma(-\nu-1/2, z_j t_N) & \text{otherwise .} \end{cases} \tag{C.13}$$

Fortunately the Gamma functions are available in the algebraic manipulation programme MAPLE, so we are finished with this part of (C.12). Concerning the calculation of the finite integral we have to take into account that $F_2(t)$ strongly oscillates with a frequency proportional to $1/\kappa$. From several tests we found that in this case the accurracy of the numerical integration routine implemented in MAPLE is not sufficient unless one subdivides the integration intervall in such a way that each part roughly contains one period of the function $F_2(t)$. This would mean to solve approximately $10/\kappa$ integrals numerically taking a disproportionately large amount of computer time. Therefore, instead of a numerical integration routine we use the series representation [20, sect. 8.440] for the Bessel functions

$$\prod_{k=1}^{3} J_{r_k}(t) = \left(\frac{t}{2}\right)^{r_2+r_2+r_3} \sum_{n=0}^{\infty} \xi_n \left(\frac{t}{2}\right)^{2n} \tag{C.14}$$

with some coefficients $\xi_n$. Since this series is convergent it can be approximated by truncation. If we demand that the error is less than $\varepsilon$ the order of truncation is about 100 which is certainly no problem if one has an algebraic manipulation programme at hand. The remaining integrations can be done analytically, so there is no further loss of accurracy.

In case one is only interested in the small momentum behaviour of the Green's function it suffices to expand $G_j(\boldsymbol{q};\boldsymbol{r})$ in powers of $\widehat{q}$. Starting from the representation (C.12) such a series can be easily derived. Concerning the finite integrals we simply expand the exponentials (cf. eqs. (C.6) and (C.7)) and do the $t$-integration as described above. To obtain a small momentum expansion for the infinite integrals one has to distinguish between two cases. If $z_j(\widehat{q}=0) \neq 0$ we can make use of the lower line of eq. (C.13). According to [20, sect. 8.354] the incomplete Gamma functions have the series representation

$$\begin{aligned} \Gamma(-\nu-1/2, z_j t_N) &= \Gamma(-\nu-1/2, x_j) \\ &\quad - \mathrm{e}^{-x_j}\, x_j^{-\nu-3/2} \sum_{n=0}^{\infty} \frac{(-1)^n \Gamma(\nu+n+3/2)}{n!\, \Gamma(\nu+3/2)}\, \mathrm{e}^{-y_j} \sum_{l=n+1}^{\infty} \frac{y_j^l}{l!} , \end{aligned} \tag{C.15}$$

where

$$y_j = (-1)^j \frac{\mathrm{i}}{2} t_N \widehat{q}^2 , \quad x_j = z_j - y_j . \tag{C.16}$$

However, for $\mu = 3$ it follows that $z_1 = -(\mathrm{i}/2)\widehat{q}^2$ leading to the contributions

$$-\frac{\mathrm{i}}{2} \sum_{\nu=0}^{\infty} b_{3,\nu}(\boldsymbol{r}) \underbrace{\int_{t_N}^{\infty} \mathrm{d}t \exp\left\{(\mathrm{i}/2)t\widehat{q}^2\right\} t^{-\nu-3/2}}_{I_{\nu+1}} . \tag{C.17}$$

These integrals cannot be represented through incomplete Gamma functions if $\widehat{q} = 0$ (cf. the upper line of eq. C.13). Therefore the expansion around $\widehat{q} = 0$ must be done differently. By partial integration one deduces a recurrence relation

$$I_{\nu+1} = (\nu+1/2)^{-1} \left( \exp\left\{(\mathrm{i}/2)t\widehat{q}^2\right\} t^{-\nu-1/2} + \frac{\mathrm{i}}{2} \widehat{q}^2 I_\nu \right) . \tag{C.18}$$

The first contribution can be expanded trivially whereas the integral is iterated again. What remains is the integral

$$I_0 = \int_{t_N}^{\infty} \mathrm{d}t \exp\left\{(\mathrm{i}/2)t\widehat{q}^2\right\} t^{-1/2} . \tag{C.19}$$

Table 2: Results for $G_\nu(0)$.

| $\nu$ | $\kappa = 0.01$ | $\kappa = 0.05$ | $\kappa = 0.1$ |
|---|---|---|---|
| 0 | $-0.1238..$ | $-0.1149$ | $-0.1055..$ |
| 1 | $-0.0386..\,\mathrm{i}$ | $-0.0345..\,\mathrm{i}$ | $-0.0306..\,\mathrm{i}$ |
| 2 | $-0.0065..$ | $-0.0078..$ | $-0.0089..$ |
| 3 | $-0.0050..\,\mathrm{i}$ | $-0.0050..\,\mathrm{i}$ | $-0.0050..\,\mathrm{i}$ |
| 4 | $0.0075 + 0.0050..\,\mathrm{i}$ | $0.0066 + 0.0045..\,\mathrm{i}$ | $0.0056.. + 0.0040..\,\mathrm{i}$ |
| 5 | $-0.00068..\,\mathrm{i}$ | $-0.00070..\mathrm{i}$ | $-0.00071..\,\mathrm{i}$ |

Table 3: Results for $G_\nu(\widehat{1})$.

| $\nu$ | $\kappa = 0.01$ | $\kappa = 0.05$ | $\kappa = 0.1$ |
|---|---|---|---|
| 0 | $-0.0417..$ | $-0.0373..$ | $-0.0329..$ |
| 1 | $-0.0386..\,\mathrm{i}$ | $-0.0345..\,\mathrm{i}$ | $-0.0306..\,\mathrm{i}$ |
| 2 | $0.0143..$ | $0.0122..$ | $0.0102..$ |
| 3 | $0.0014..\,\mathrm{i}$ | $0.00068..\,\mathrm{i}$ | $0.00009..\,\mathrm{i}$ |
| 4 | $-0.0065 + 0.0024..\,\mathrm{i}$ | $-0.0051 + 0.0022..\,\mathrm{i}$ | $-0.0039.. + 0.0019..\,\mathrm{i}$ |
| 5 | $0.00014..\,\mathrm{i}$ | $0.00014..\mathrm{i}$ | $0.00011..\,\mathrm{i}$ |

If we shift the lower limit of integration to zero we can use [20, sect. 3.691] and obtain

$$I_0 = \frac{\sqrt{\mathrm{i}\pi}}{\widehat{q}}(1-\mathrm{i}) - \int_0^{t_N} \mathrm{d}t \exp\left\{(\mathrm{i}/2)t\widehat{q}^2\right\} t^{-1/2} \ . \tag{C.20}$$

Collecting the results one ends up with the desired small momentum expansion for the Green's function

$$G(\boldsymbol{q};\boldsymbol{r}) = \sum_{\nu=0}^{\infty} G_\nu(\boldsymbol{r})\widehat{q}^\nu \ . \tag{C.21}$$

The first five coefficients for $\boldsymbol{r} = 0$, $\boldsymbol{r} = \widehat{1}$ and some values of the inverse coupling $\kappa$ are listed in table 2 and 3.